\documentclass[preprint,12pt,nofootinbib]{revtex4}

\usepackage{color}
\usepackage[brazil, english]{babel}
\usepackage[utf8]{inputenc}
\usepackage{graphicx}
\usepackage{epsfig}
\usepackage{amssymb}
\usepackage{amsmath} 
\usepackage{slashed}
\usepackage[unicode=true,bookmarks=false,breaklinks=false,pdfborder={0 0 1},colorlinks=true]
 {hyperref}
\hypersetup{
 citecolor=blue,linkcolor=blue,urlcolor=blue}

\graphicspath{%
    {converted_graphics/}
    {/}
}
\begin{document}


\title{Fermionic and bosonic fluctuation-dissipation theorem \\ from a deformed AdS holographic model \\ at finite temperature and chemical potential} 

\author{Nathan G. Caldeira$^{1}$}
\email[Eletronic address: ]{nathangomesc@hotmail.com}
\author{Eduardo Folco Capossoli$^{2}$}
\email[Eletronic address: ]{eduardo\_capossoli@cp2.g12.br}
\author{Carlos A. D. Zarro$^{1}$}
\email[Eletronic address: ]{carlos.zarro@if.ufrj.br}
\author{Henrique Boschi-Filho$^{1}$}
\email[Eletronic address: ]{boschi@if.ufrj.br}  
\affiliation{$^1$Instituto de F\'{\i}sica, Universidade Federal do Rio de Janeiro, 21.941-972 - Rio de Janeiro-RJ - Brazil \\ 
 $^2$Departamento de F\'{\i}sica and Mestrado Profissional em Práticas de Educa\c{c}\~{a}o B\'{a}sica (MPPEB), 
 Col\'egio Pedro II, 20.921-903 - Rio de Janeiro-RJ - Brazil
}

\begin{abstract}
In this work we study fluctuations and dissipation of a string in a deformed anti-de Sitter (AdS) space at finite temperature and density. The deformed AdS space is a charged black hole solution of the Einstein-Maxwell-Dilaton action. In this background we take into account the backreaction on the horizon function from an exponential deformation of the AdS space.
From this model we compute the admittance and study the influence of the temperature and the  chemical potential on it. We calculate the two-point correlations functions, and the mean square displacement for bosonic and fermionic cases, from which we obtain the short and large time approximations. For the long time, we obtain a sub-diffusive regime $\sim \log t$. Combining the results from the admittance and the correlations functions we check the fluctuation-dissipation theorem for bosonic and fermionic systems. 
\end{abstract}


\maketitle


\section{Introduction}

Fluctuation and dissipation are ubiquitous phenomena in statistical physics. These two quantities are related by the famous fluctuation-dissipation theorem 
\cite{Kubo1966, Kubo1991}. 
The natural set up to observe these properties is a system at finite temperature and density. In particular, ensembles at finite temperature combined with finite particle density captures a lot of  attention from the high-energy community, specially for both high temperature and chemical potential. In this extreme environment, nuclear matter is totally deconfined, and is known as the quark-gluon plasma (QGP) \cite{Collins:1974ky, Cabibbo:1975ig, Shuryak:2014zxa}. By analyzing it, one can study from hadronic constituents to the interior of neutron stars. Many important experiments were idealized in order to reproduce on Earth these conditions as can be seen at Relativistic Heavy Ion Collider (RHIC) \cite{Arsene:2004fa, Back:2004je, Adams:2005dq, Adcox:2004mh} and ALICE at CERN \cite{Aamodt:2008zz, Aamodt:2010cz}. One can reasonably argue \cite{Moore:2020wvy} that QGP can be understood within two different points of view: the thermodynamics and transport properties. Regarding the transport properties one can particularize their attention, for instance, on thermal fluctuations. An interesting review which relates QGP and fluctuations is presented in Ref. \cite{Bluhm:2020mpc}.

Concerned, in particular, with finite and high chemical potential, as we are dealing with an extreme environment, the calculations are very hard and usually it are not trustworthy. This means that lattice QCD starts to crash at baryonic chemical potential $\mu_B/T \gtrsim 1$ \cite{ Kadam:2020utt}. Such difficulties are usually related to an intrinsic characteristic of the canonical theoretical formulation known as the sign problem \cite{Goy:2016egl, Troyer:2004ge}. A recent and important progress in this direction has been done where an  imaginary baryonic density was extrapolated to a real chemical potential around $\mu\sim 300$ MeV  \cite{Borsanyi:2020fev}.

Among many proposals to tackle QCD, or even lattice QCD limitations, one can resort to the AdS/CFT correspondence, or in a broader sense  string/gauge duality, originally proposed by Maldacena \cite{Aharony:1999ti}. In one of its forms, it relates a weak coupling theory living in five-dimensional anti-de Sitter (AdS) space to a conformal $SU(N_c \to \infty)$ ${\cal N}=4$ SYM QCD-like theory living in a four-dimensional Minkowski space. Mathematically speaking,  such a duality can be stated since in supergravity (SUGRA) approximation of string theory in the AdS space, both theories can be related through $Z_{\rm CFT}[\varphi_o]= \langle   \exp  ( \int_{\partial \Omega} d^4 x \; {\cal O} \varphi_o  ) \rangle  = \int_{\varphi_o} D \varphi \exp (-I_s(\varphi))$ with $\varphi$ representing a non-normalizable SUGRA field, $ I_s(\varphi)$ is the corresponding on shell SUGRA action, $\varphi_o$ is the value of $\varphi$ at the boundary $\partial \Omega$, and $\cal O$ is the associated operator of the conformal field theory (CFT). 
Making use of this correspondence many works were done in order to study hot dense QGP \cite{Policastro:2001yc, Kim:2007xi, Panero:2009tv, DeWolfe:2013cua, CasalderreySolana:2011us, Arefeva:2020vae}, considering finite chemical potentials (or some related topics as for instance, QCD phase transition, chiral symmetry breaking, critical exponents, etc. \cite{Colangelo:2010pe, Li:2012ay, Bohra:2019ebj, Ghoroku:2020fkv, Evans:2020whc, Cao:2020ske, Cao:2020ryx, He:2020fdi, Ballon-Bayona:2020xls, Mamani:2020pks, Braga:2019xwl, Braga:2020myi, Rodrigues:2020ndy, Chen:2020ath, Ballon-Bayona:2020xtf}) and quantum or thermal fluctuations such as the Brownian motion \cite{deBoer:2008gu, Atmaja:2010uu, Chakrabortty:2013kra, Sadeghi:2013jja, Banerjee:2013rca, Banerjee:2015vmo, Chakrabarty:2019aeu}, fluctuation and dissipation \cite{Tong:2012nf, Edalati:2012tc, Kiritsis:2012ta, Fischler:2014tka, Roychowdhury:2015mta, Banerjee:2015fed, Giataganas:2018ekx, Caldeira:2020sot, Caldeira:2020rir}, drag forces \cite{Gubser:2006bz, Gubser:2006qh, Andreev:2017bvr, Andreev:2018emc, Diles:2019jkw}, or related topics  \cite{Kinar:1999xu, Gursoy:2010aa, Giataganas:2013zaa, Giataganas:2013hwa, Dudal:2014jfa, Dudal:2018rki}.


Here in this work we will focus on the AdS/CFT correspondence to study holographically thermal fluctuations and dissipation modeled by a probe string attached to a probe brane in a  deformed AdS-Reissner-Nördstrom (AdS-RN) spacetime with backreaction. 
 Charged black holes, in the AdS/CFT scenario, were considered in the very first time in the Refs. \cite{Chamblin:1999tk, Chamblin:1999hg}.
The starting point of our model is an Einstein-Maxwell-Dilaton (EMD) string/gauge model at finite temperature and chemical potential. The holographic  EMD and  Einstein-Maxwell (EM) formulations, are well-known ways to represent non-conformal plasmas and were considered in many works. An incomplete list, or even already cited throughout the text, can be seen in Refs. \cite{Li:2011hp, Cai:2012xh, He:2013qq, Li:2014hja, Yang:2014bqa, Li:2017ple, Chen:2018vty, Arefeva:2018hyo, Chen:2019rez,  Arefeva:2018cli, Colangelo:2020tpr, Bohra:2020qom}.

Regarding the deformation in our AdS-RN spacetime we are considering, it can be understood as the introduction of a conformal exponential factor $\zeta(z)$, as a function of the holographic coordinate $z$, in the metric, written in the following form  \cite{Ballon-Bayona:2017sxa, Ballon-Bayona:2020xls, Caldeira:2020rir}:
\begin{eqnarray}
ds^2 &=& \dfrac{L^2}{\zeta(z)^2}\left(\dfrac{dz^2}{f(z)} - f(z)dt^2 + d\vec{x}^2\right),\,\,\,{\rm with} \label{Metriczetanovo} \\
\zeta(z) &=& z \, e^{-\frac{1}{2} \left(k z^2\right)}, \label{warpnovo}
\end{eqnarray}
\noindent where $k$ is a constant to be fixed latter and the AdS radius will be set as $L=1$. Notice that the introduction of this deformation was proposed in Refs. \cite{Andreev:2006vy, Andreev:2006ct} as an alternative to the famous Softwall model (SWM) in Ref. \cite{Karch:2006pv}, to  break the conformal symmetry in order to generate a confining potential for a quark-antiquark pair. Besides, 
such a deformation in the AdS metric was successfully applied in different problems within string/gauge duality, as can be seen, for instance in Refs. \cite{ Afonin:2012jn, Rinaldi:2017wdn, Bruni:2018dqm, Diles:2018wbe, Afonin:2018era,  FolcoCapossoli:2019imm, Rinaldi:2020ssz, Caldeira:2020sot, FolcoCapossoli:2020pks, Caldeira:2020rir,Contreras:2021onc, Contreras:2021epz}. If the limits $k\to 0$ and $f(z)\to 1$ are taken, one recovers the  AdS spacetime in Poincaré coordinates.

It is worthwhile to mention that in this work the horizon function $f(z)$ will be properly achieved from the solutions of the EMD geometric background. 
This horizon function represents  a deformed and backreacted  AdS-Reissner-N\"ordstrom (AdS-RN) spacetime.  In this sense we will be taking into account  backreaction contributions from the metric deformation and considering the string and the brane in the probe approximation.  

 Since we established the basic set up of our work, the reader will see it as in the following: In section \ref{sec2}, we state the fluctuation-dissipation theorem for bosons and fermions.  In section \ref{geo}, we present the mathematical description of the EMD geometry as well as the computation of our backreacted horizon function with finite chemical potential and the corresponding Hawking temperature. In Section \ref{adcor}, from the Nambu-Goto action we obtain the equations  of motion for a string endpoint in this background,    compute  the admittance, the two point correlation functions, and the mean square displacement. For this last quantity we compute separately the bosonic ($\mu<0$) and fermionic ($\mu >0$) cases from which we obtain the short and long time behaviors. In section \ref{sec5}, we check the fluctuation-dissipation theorem in this set up for bosons and fermions. Finally, in Sec. \ref{conc} we present our conclusions.

\section{Fluctuation-dissipation theorem}
\label{sec2}

In this section we are going to state the fluctuation-dissipation theorem, for the bosonic and fermionic cases which depend crucially on the sign of the chemical potential $\mu$.   

First of all, let us define the symmetric Green's function in terms of the thermal correlation functions as:
\begin{equation}
 G_{\rm Sym}(t)\equiv \frac{1}{2}\left(\langle x(t)x(0) \rangle+\langle x(0)x(t) \rangle\right)\,. 
\end{equation}

For the bosonic case ($\mu < 0$), we have that the fluctuation-dissipation theorem  in the presence of a chemical potential can be written as \cite{Zubarev}
\begin{eqnarray}
\label{FDTB}
    G_{\rm Sym}(t) =\mathcal{F}^{-1}\left[\left(1+2n_{B}\right)\Im \chi (\omega)\right]
    \,,
\end{eqnarray}
and for the fermionic case ($\mu > 0$)one has  \cite{Markov:2009ue}
\begin{eqnarray}
\label{FDTF}
    G_{\rm Sym}(t)
    =\mathcal{F}^{-1}\left[\left(1+2n_{F}\right)\Im \chi (\omega)\right]
    \,, 
\end{eqnarray}
where $\mathcal{F}^{-1}[\cdots]$ is the inverse Fourier transform, $n_{B}$ is the Bose-Einstein distribution, $n_{F}$ the Fermi-Dirac distribution, and $\Im \chi(\omega)$ is the imaginary part of the admittance. 

In the following, we are going to define our holographic set up of a fluctuating string in a deformed AdS background with finite temperature and chemical potential from where we obtain the admittance and the correlation functions. As far as we are concerned, we could not find a discussion of the fluctuation-dissipation theorem in the holographic approach for bosons and fermions at finite density.

\section{The  deformed and backreacted  black hole} \label{geo}

Here, we will discuss our gravitational background starting from the EMD system. It was used in many references to deal with finite chemical potential. The action for this theory is written in the Einstein frame as:
\begin{equation} \label{EMD action}
S = \dfrac{1}{16\pi G_5}\int d^{5}x \sqrt{-g}\left(R - \dfrac{4}{3}g^{mn}\partial_{m}\phi\partial_{n}\phi + V(\phi)-\dfrac{1}{4}F_{mn}F^{mn} \right),
\end{equation}
where $ G_5 $ is the 5-dimensional Newton's constant, $g$ is the metric determinant, $ R $ is the Ricci scalar, $F_{mn} = \partial_{m} A_{n} - \partial_{n} A_{m}$ is the Maxwell field, $\phi$ is the dilaton field,  $V(\phi)$ its potential and $m,n=0,1,2,3,5$.

Thus, from this action one can obtain the following field equations:
\begin{eqnarray}
G_{mn} -\dfrac{4}{3}\left(\partial_{m}\phi\partial_{n}\phi - \dfrac{1}{2}g_{mn}(\partial \phi)^2\right) - \dfrac{1}{2}g_{mn}V(\phi) - \frac{1}{2}\left(F_{ma}{F_{n}}^{a} - \dfrac{1}{4}g_{mn}F^2\right) &=& 0, \label{EinsteinEqn} \label{eom1}\qquad \\ 
\Box \phi + \dfrac{3}{8}\frac{\partial\,V(\phi )}{\partial\,\phi}&=& 0, \label{DilatonEqn} \label{eom2}\\
\nabla_{m}F^{mn} &=& 0, \label{eom3}
\end{eqnarray}
\noindent where the Einstein tensor $ G_{mn} $ can be written as
\begin{equation}
G_{mn} = R_{mn}- \dfrac{1}{2}g_{mn}R.
\end{equation}
In order to solve Eqs. \eqref{eom1}, \eqref{eom2} and \eqref{eom3} we will used the metric given by Eq. \eqref{Metriczetanovo}, with the warp factor given by Eq. \eqref{warpnovo}. Furthermore we will choose 
\begin{eqnarray}
A_{m}(z) &=& (0, A_{t}(z), {\vec{0}})\,,  \label{At}
\end{eqnarray}
\noindent where $A_t(z)$ is the time component of the $U(1)$ gauge field $A_{m}(z)$ which is  dual to the presence of a global conserved current in the $4D$ dual theory.

Replacing Eqs. \eqref{Metriczetanovo} and  \eqref{At} into Eqs.  \eqref{eom1} and \eqref{eom3} one gets the set of coupled equations: 
\begin{eqnarray}
\frac{\zeta''(z)}{\zeta(z)} - \frac{4}{9}\phi'(z)^2 &=& 0,\label{breqn3} \\
\frac{\zeta'(z)}{\zeta(z)} - \frac{A_{t}''(z)}{A_{t}'(z)} &=& 0,\label{Atreqn3} \\
\frac{d}{dz}\left(\zeta(z)^{-3}f'(z)\right) - \frac{A_{t}'(z)^2}{\zeta(z)} &=& 0,\label{freqn3} 
\end{eqnarray}
where $'$ represents derivative with respect to $z$. Solving Eq.  \eqref{breqn3}, one obtains the dilaton field:
\begin{eqnarray}
    \phi(z) = 
    \pm \frac{3}{4}\left( \sqrt{k \left(k z^2-3\right)}z-3 \log \left( \frac{\sqrt{k\left(k z^2-3\right)
    }+k z}{C}\right)\right)\,,
\end{eqnarray}
where $C$ is a constant with units of energy.
By using Eq. \eqref{eom2}, one gets the dilaton potential:
\begin{eqnarray}  \label{potential}
V(\phi) = 12\zeta'(z)^2 f(z) - 3\zeta'(z)f'(z)\zeta(z) - \frac{4}{3}f(z)\zeta (z)^2\phi'(z)^2 + \frac{1}{2}\zeta(z)^{4} A_{t}'(z)^2  \;\;.
\end{eqnarray}

Plugging the warp factor from Eq. \eqref{warpnovo} in Eq. \eqref{Atreqn3} we get the following general solution for $A_t(z)$:
\begin{equation}
    A_{t}(z)= -\frac{C_1 e^{-\frac{1}{2} \left(k z^2\right)}}{k} + C_2\,,
\end{equation}
\noindent where $C_1$ and $C_2$ are constants to be fixed by imposing the regularity conditions at the horizon. 
These conditions are $ A_{t}(z=z_h) = 0$ and  $A_{t}(z=0) = \mu$, where $ \mu $ is the chemical potential of the dual gauge theory. 
Hence, after some algebra one gets the expression for $A_t(z)$, given by:
\begin{equation}\label{at}
   A_t(z) = \mu \left(\frac{e^{\frac{1}{2} k \left(z_h^2-z^2\right)}-1}{e^{\frac{k }{2}z_h^2}-1}\right).
\end{equation}

If one wants to recover the AdS-RN space, one can take the limit $k \to 0$, so that:
\begin{equation}
    A_t^{\rm AdS-RN}(z) = \mu  \left(1-\frac{z^2}{z_h^2}\right).
\end{equation}

\begin{figure}[!ht] 
\vskip 0.5cm
	\centering
	\includegraphics[scale = 0.30]{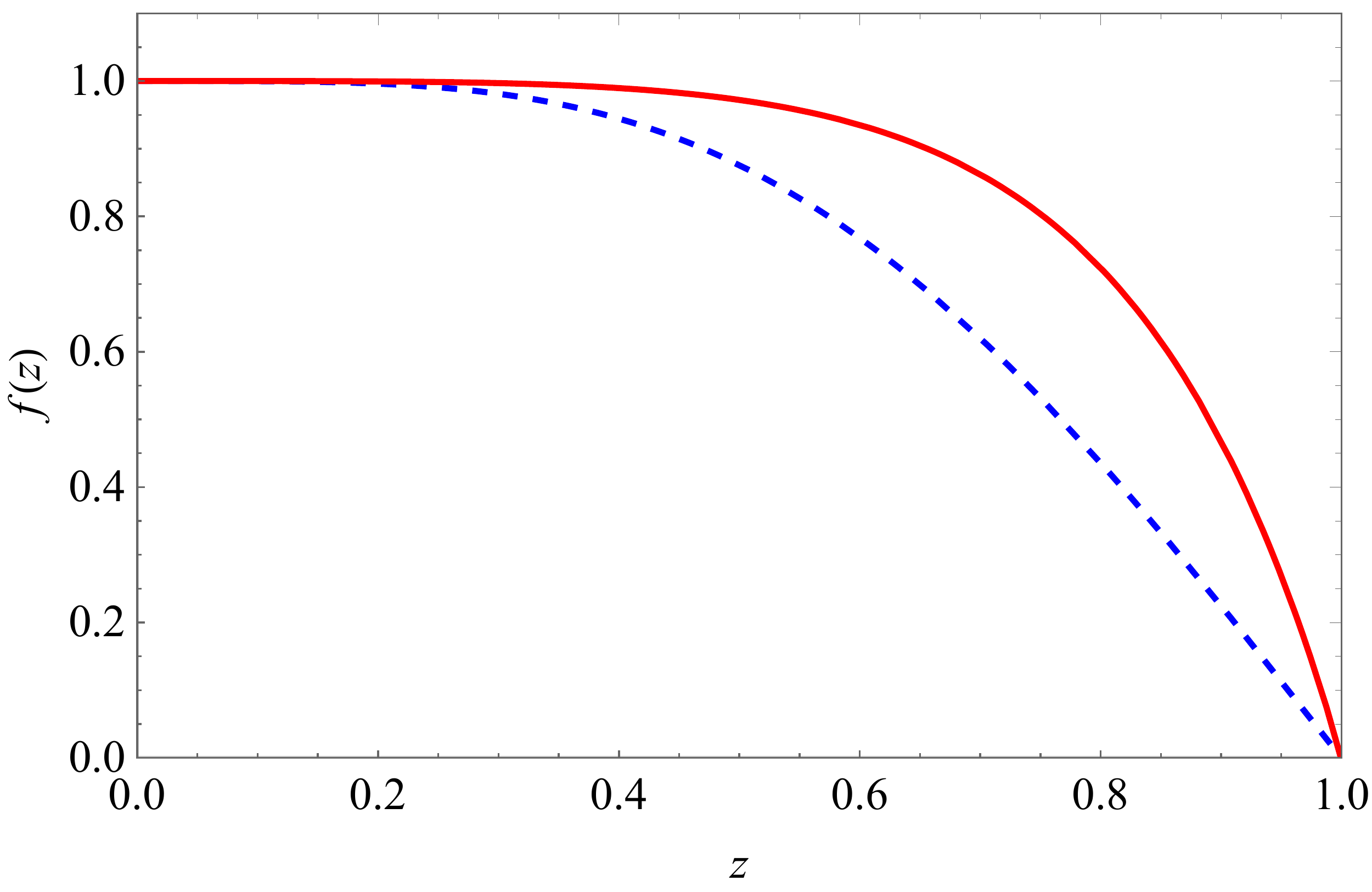}
	\includegraphics[scale = 0.40]{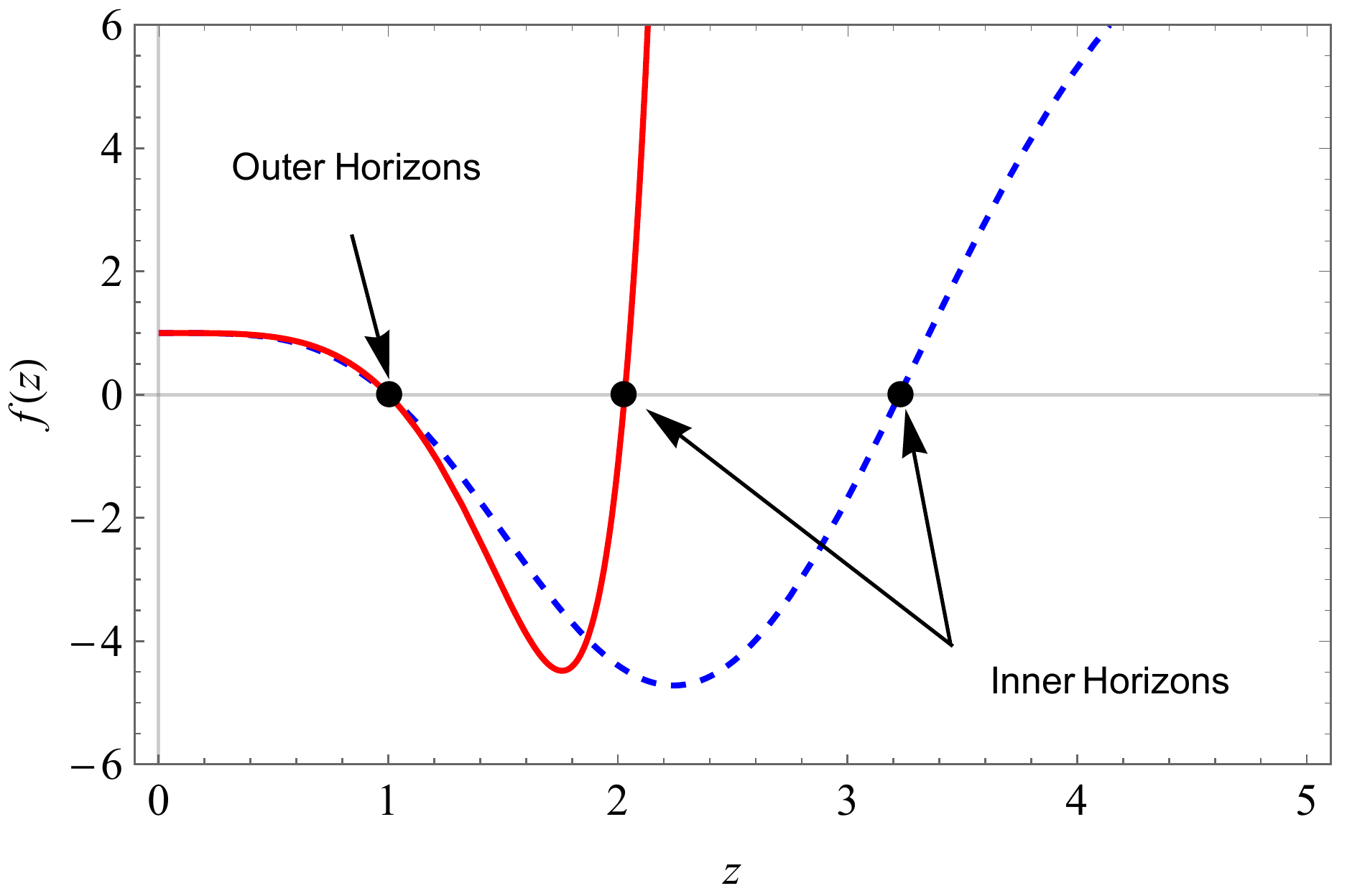}
	\vskip -0.5cm
	\caption{{\sl Left panel:}  Horizon function as a function o $z$ for $\mu=0$. The dashed blue curve represents $k=1$ and the red curve represents $k= -1$. {\sl Right panel:} Horizon function as a function of $z$ for $\mu=0.8$. The dashed blue curve represents $k=0.18$ and the red curve represents $k= -0.18$.}
	\label{edu}
\end{figure}

Now, using Eqs. \eqref{warpnovo} and \eqref{Atreqn3}, satisfying  $f(0) = 1$ (pure AdS ) and the horizon property $f(z_h) = 0$, one can solve analytically Eq. \eqref{freqn3}, so that: 
\begin{align}\label{horizon}
f(z)&=1-\frac{{\cal G}(z)}{{\cal G}(z_h)}
e^{\frac{3k}{2}(  z_{h} ^2 -  z^2)}
-\frac{\mu^{2}\, {\cal H}(z_h)}{72 k \left(e^{\frac{1}{2}k z_{h} ^2}-1\right)^2}
\Bigg[\frac{{\cal G}(z)}{{\cal G}(z_h)}
\, e^{\frac{k}{2}(  z_{h} ^2 -3  z^2)}-\frac{{\cal H}(z)}{{\cal H}(z_h)}{ e^{k \left(z_{h} ^2-2 z^2\right)}}\Bigg]\,,
\end{align}
where
\begin{align}
{{\cal G}(z)}\equiv 3 k z^2-2 e^{\frac{3}{2} k z^2}+2\;; 
\qquad
{\cal H}(z)\equiv 18 k z^2+7 e^{2 k z^2}-24 e^{\frac{1}{2}k z^2} k z^2-16e^{\frac{1}{2}k z^2}+9\,. 
\end{align}
In Figure \ref{edu} we present the behavior of the horizon function, coming from Eq. \eqref{horizon}, in terms of the holographic coordinate $z$. Note that in all Figures of this work we use arbitrary units for the physical parameters $k$, $\mu$, etc. 
Moreover, one can verify that horizon function $f(z)$ for the AdS-RN space is recovered for $k \to 0$ which is given by:
\begin{equation}
  f^{\rm AdS-RN}(z) =  1 -\frac{z^4}{z_h^4}-\frac{\mu ^2z_h^2}{3} \left(\frac{z^4}{ z_h^4}-\frac{z^6}{ z_h^6}\right)\,, 
\end{equation}
in agreement with the literature, see e. g., \cite{Hartnoll:2009sz}.

\subsection{Hawking temperature}\label{HT}

In this subsection we will calculate the Hawking temperature associated with the black hole present in our deformed and backreacted AdS-RN spacetime which is the temperature of the bath at the boundary. Firstly, let us make a change of coordinates $r = 1/z$. In this coordinate system the horizon function Eq. \eqref{horizon}
 reads: 
\begin{align}\label{horizonr}
f(r)&=1-\frac{{\cal G}(r)}{{\cal G}(r_h)}
e^{\frac{3k}{2}(  r_{h} ^{-2} -  r^{-2})}
-\frac{\mu^{2}\, {\cal H}(r_h)}{72 k \left(e^{\frac{1}{2}k r_{h} ^{-2}}-1\right)^2}
\Bigg[\frac{{\cal G}(r)}{{\cal G}(r_h)}
\, e^{\frac{k}{2}(  r_{h} ^{-2} -3  r^{-2})}-\frac{{\cal H}(r)}{{\cal H}(r_h)}{ e^{k \left(r_{h} ^{-2}-2 r^{-2}\right)}}\Bigg]
\end{align}
where
\begin{align}
{{\cal G}(r)}\equiv 3 k r^{-2}-2 e^{\frac{3}{2} k r^{-2}}+2\;; 
\qquad
{\cal H}(r)\equiv 18 k r^{-2}+7 e^{2 k r^{-2}}-24 e^{\frac{1}{2}k r^{-2}} k r^{-2}-16e^{\frac{1}{2}k r^{-2}}+9\,. 
\end{align}

The Hawking temperature can be obtained from Eq. \eqref{horizonr}, so that:
\begin{eqnarray}
 \label{HawkingTemperature}
 T = \left|\frac{r^{2}}{4\pi}\frac{d f(r)}{d r}\right|\Bigg|_{r=r_{h}}
 =\frac{r_{h}}{\pi}\left|g(x)-\frac{\mu^2}{k}c(x)\right|,
\end{eqnarray}
where $x\equiv {k}/{r^2_{h}}$, 
\begin{equation}
  g(x) \equiv \frac{9 x^2}{4\left(2 \left( e^{\frac{3}{2}x}-1\right
 )-3x \right)} 
\end{equation}
\noindent and 
\begin{equation}
\label{cdefinition}
    c(x) \equiv \frac{ e^{-\frac{3 x}{2}} \left(6 e^{x/2} x^3+7 e^{x/2} x^2+9 e^{\frac{5 x}{2}} x^2-16 e^{2 x} x^2\right)}{32 \left(e^{x/2}-1\right)^2 \left(-3 x+2 e^{\frac{3 x}{2}}-2\right)}\,.
\end{equation}

\begin{figure}[!ht]
	\centering
	\includegraphics[scale = 0.3]{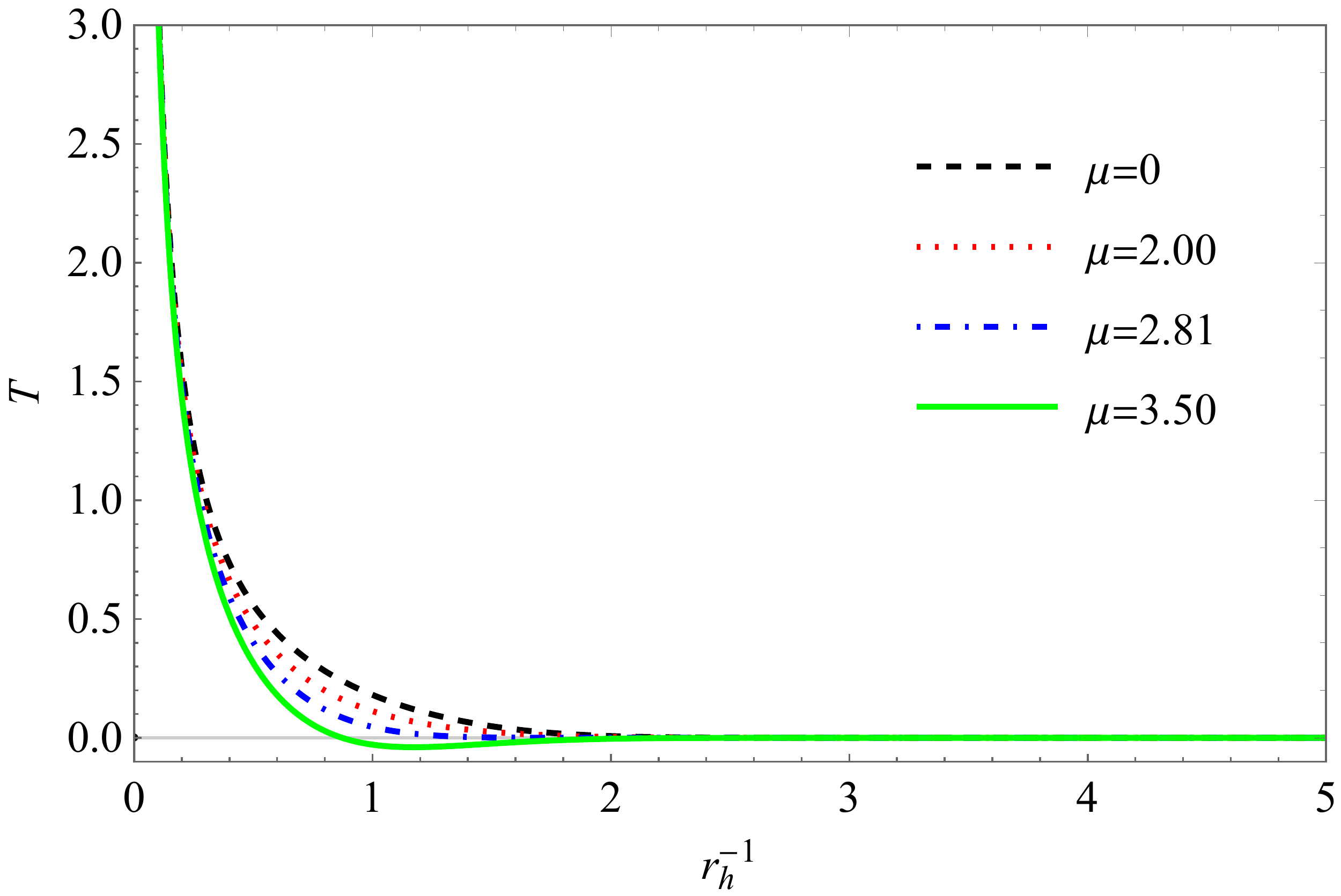}
	\includegraphics[scale = 0.3]{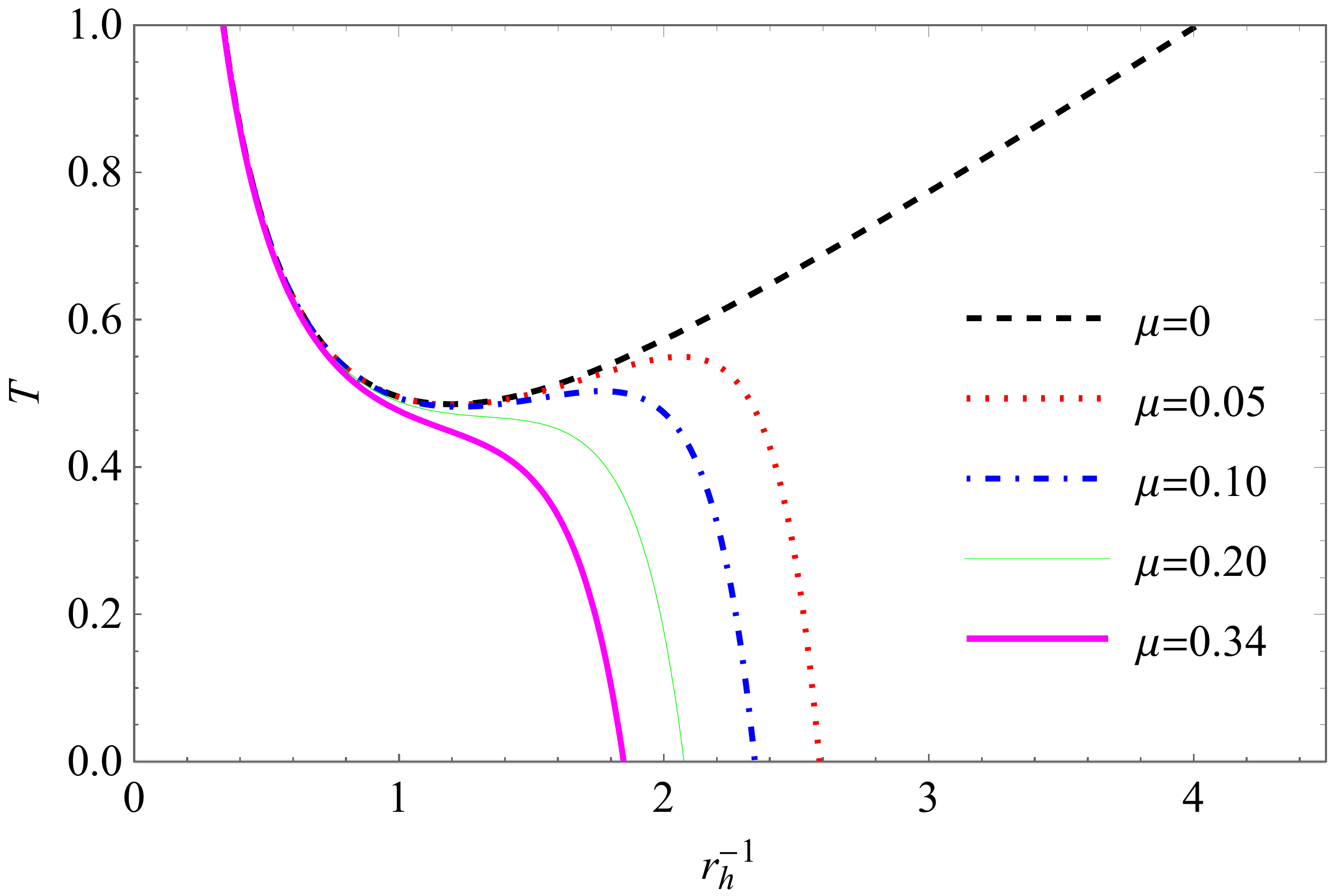}
	\caption{{\sl Left panel:} Temperature versus $r_h^{-1}$ for a fixed $k=1 $ and some values of $\mu$. The value  $\mu = 2.81$ corresponds to a critical chemical potential. For $\mu> \mu_c$ the temperature would become negative. {\sl Right panel:} Temperature versus $r_h^{-1}$ for a fixed $k=-1 $ and some values of $\mu$. The critical chemical potential is $\mu_c = 0.34$. For $\mu > \mu_c$ the temperature only decreases.}
	\label{mukpos}
\end{figure}

Then, the allowed physical values for $\mu$ are 
\begin{eqnarray}
-\sqrt{\frac{72 e^x \left(e^{x/2}-1\right)^2 x}{6 x-16 e^{\frac{3 x}{2}}+9 e^{2 x}+7}} &\leq& \frac{\mu}{r_h} \leq \sqrt{\frac{72 e^x \left(e^{x/2}-1\right)^2 x}{6 x-16 e^{\frac{3 x}{2}}+9 e^{2 x}+7}}.
\end{eqnarray}
This expression represents  constraints  for the chemical potential $\mu$ and the constant $k$. It's worthwhile to mention that for $k \to 0$ one recovers 
\begin{equation}
  - \sqrt{6} \leq \frac{\mu}{r_h} \leq \sqrt{6}, 
\end{equation}
\noindent as expected for AdS-RN black hole, as one can see in Ref. \cite{Hartnoll:2009sz}, where the author considered $\mu \ge 0$. 
In Figure \ref{mukpos}, we present the behavior of Hawking temperature as a function of $r_h$ for positive and negative values of $k$, with different values of the chemical potential. 

\section{Admittance, correlation functions and  mean square displacement} \label{adcor}

In this section we will obtain the equations of motion from Nambu-Goto action and compute the main quantities related to the Brownian motion, such as, the admittance, the mean square displacement and the correlation functions.

\subsection{Nambu-Goto action and equations of motion}\label{ngsec}

Here, we will consider our probe string attached to a probe brane in the EMD background. This system is dual to a particle in a thermal bath with a finite chemical potential.  The metric of the deformed and backreacted AdS-RN spacetime Eq. \eqref{Metriczetanovo}, in the $r$ coordinate system, is  written as:
\begin{equation}\label{metrictemp}
    ds^2 = e^{\frac{k}{r^2}} \left[-r^{2}f(r)dt^{2}+{r^2}\left(\eta_{i j} dx^{i}dx^{j} \right) +\frac{dr^2}{r^{2}f(r)}\right].
\end{equation}
Besides, the dynamics of such a string will be given by the Nambu-Goto action:
\begin{equation}\label{ng}
S_{NG} = - \frac{1}{2 \pi \alpha'} \int d\tau d\sigma \sqrt{-\gamma}\, ,
\end{equation}
\noindent where $\alpha'$ is the Regge slope, $\gamma = {\rm det} (\gamma_{\alpha \beta})$ and $\gamma_{\alpha \beta} = g_{mn} \partial_{\alpha}X^m \partial_{\beta}X^n $ is the induced metric on the worldsheet with $m,n = 0, 1, 2, 3, 5$.

 We choose a static gauge, where $t = \tau$, $r = \sigma$ and $X= X(\tau, \sigma)$, as done in Refs. \cite{Edalati:2012tc, Giataganas:2018ekx}. By using the metric, Eq. \eqref{metrictemp}, and expanding the Nambu-Goto action, Eq. \eqref{ng}, in order to keep the quadratic terms $\dot{X}^2$, $X'^2$, we get: 
\begin{eqnarray}\label{ngapprox}
S_{NG} \approx
- \frac{1}{4 \pi \alpha'} \int d\tau d\sigma \left[ \;\dot{X}^2 \frac{e^{\frac{ k}{r^2}}}{f(r)}-X'^2 r^4 f(r) e^{\frac{k}{r^2}} \right]\,,
\end{eqnarray}
\noindent where $\dot{X}=\partial_{\tau =t} X$ and $X'=\partial_{\sigma =r} X$.

The equation of motion is
\begin{equation}
    \frac{\partial }{\partial r}\left(r^{4}f(r)e^{\frac{k}{r^{2}}}X'(t,r)\right)-\frac{e^{\frac{k}{r^{2}}}}{f(r)}\ddot{X}(t,r)=0\,, 
\end{equation}
\noindent where $f(r)$ is given by Eq. \eqref{horizonr} which  can be expanded close to the horizon ($r \sim r_h$) as $f(r)\approx f'(r_h) (r-r_{h})$.
Using the ansatz $X(t,r)=e^{i\omega t}h_{\omega}(r)$, one gets: 
\begin{equation}
\label{ansatz}
    r^4  f(r) h''(r)+\left[-2 k r  f(r)+r^4  f'(r)+4 r^3  f(r)\right] h'(r)+\frac{\omega ^2  }{f(r)}h(r)=0\,.
\end{equation}
Changing to a tortoise coordinate  
   $  r_{*}=\int {dr}({r^{2}f(r)})^{-1}\,,$ 
we get
\begin{eqnarray}
    \frac{1}{f(r)}\frac{d^{2}h_{\omega}(r_{*})}{dr^{2}_{*}}
    +\left(-\frac{2 k}{r}+2 r\right)\frac{d h_{\omega}}{dr_{*}}+\frac{\omega ^2}{f(r)}h_{\omega}(r)=0\,.
\end{eqnarray}
Making a Bogoliubov  transformation
$h_{\omega}(r_{*})=e^{B(r_*)}\psi(r_*)\,,$ 
with $B(r)= -{k}/({2 r^2})-\log (r)$, 
we obtain a Schr\"{o}dinger-like equation
\begin{equation}\label{sch}
    \frac{d^{2}\psi(r_{*})}{dr_{*}^{2}}+\left(\omega^{2}-V(r)\right)\psi(r_{*})=0,
\end{equation}
with potential 
\begin{eqnarray}
    V(r)&=&-f(r)\left[\left(-\frac{k^2}{r^2}+k-2 r^2\right) f(r)+r \left(k-r^2\right) f'(r)\right]\,.
\end{eqnarray}
From the above equation, one has $V(r=r_h) =0$, since $f(r_{h})=0$. 

The Eq. \eqref{sch} cannot be analytically solved, hence one seeks for approximate  solutions. This is discussed in Appendix \ref{apA}. The relevant solution close to the boundary is: 
\begin{eqnarray}\label{bound}
    h^{C}_{\omega}(r)\approx A_1\left(\frac{e^{-\frac{k}{2r_{h}^{2}}}}{r_{h}}
    +\frac{i \omega r_{h} e^{\frac{k}{2 r_{h} ^2}} }{3r^3}\right)\,, 
\end{eqnarray}
where $A_1$ is a normalization constant.

\subsection{Admittance}\label{A}

In this section, we will present one of the main results of this work, which is the admittance in the presence of a finite chemical potential in a backreacted geometry.  
From the solution given by Eq. \eqref{bound}, one can compute the the admittance $\chi(\omega)$, also known as the linear response, of the string endpoint on the brane. One should note that such a response is due to the action of an external force in an arbitrary brane direction, $x^{i}$, and can be represented by $F(\omega) = E \, e^{-i\omega t}$, where $E$ is the electric field on the brane.

By considering the electric field as $E = E(A_t, \vec A)$ and taking into account the  approximate Nambu-Goto action, one has:
\begin{equation}\label{eq:actionBH+E}
S \approx - \frac{1}{4 \pi \alpha'} \int d\tau d\sigma \left[ \;\dot{X}^2 \frac{e^{\frac{ k}{r^2}}}{f(r)}-X'^2 r^4 f(r) e^{\frac{k}{r^2}} \right] + \int dt \left(A_t + \vec{A} \cdot \vec{\dot{x}} \right)\Big|_{r=r_b}\,,
\end{equation}
where the second integral is a surface term. Such a term was chosen in an arbitrary direction, and it does not participate in the bulk dynamics. Choosing $A_t=0$ and integrating by parts the surface term, one gets:
\begin{equation}\label{ngbtermBH+E}
S \approx - \frac{1}{4 \pi \alpha'} \int dt dr  \left[ \;\dot{X}^2 \frac{e^{\frac{ k}{r^2}}}{f(r)}-X'^2 r^4 f(r) e^{\frac{k}{r^2}} \right] - \int dt \; F(t)\,\left.\left(\frac{\partial X(t,r)}{\partial r}\right)\right|_{r=r_{b}}, 
\end{equation}
with $\tau=t$ and $\sigma=r$. Computing $\delta S/\delta X'=0$ and imposing Neumann boundary condition on the brane, one can write the force in a Fourier domain as:
\begin{eqnarray}
 F(\omega)=\frac{A_{1}}{2 \pi \alpha'}\left[-i \omega r_{h} e^{\frac{k}{2 r_{h} ^2}} f(r_{b})e^{\frac{ k}{r_b^2}}\right]\,. 
\end{eqnarray}
More details on the calculation of $F(\omega)$ can be found with in Refs. \cite{Tong:2012nf, Edalati:2012tc, Giataganas:2018ekx, Caldeira:2020sot, Caldeira:2020rir}.

Therefore the admittance is
\begin{eqnarray}
\label{Admittance}
 \chi(\omega)\equiv \frac{h^{C}_{\omega}(\omega)}{F(\omega)}
 =\frac{2\pi i\alpha'}{ \omega r_{h}^{2} e^{\frac{k}{r_{h} ^2}}f(r_{b})e^{\frac{ k}{r_b^2}}}\left(1
 +\frac{i \omega r_{h}^{2} e^{\frac{k}{r_{h}^2}} }{r_{b}^3}\right)\,.
\end{eqnarray}
In the limit $r_{b}\gg r_{h}$, $r_{b}\gg k$ and $f(r_{b})\to 1$ the admittance reads
\begin{equation}
\label{Admittance1}
 \chi(\omega)=-2\pi\alpha'\frac{e^{-\frac{k}{r_{h}^{2}}}
 }{i \omega r_{h}^{2}}=\frac{2\pi i\alpha' e^{-x}}{\omega r_{h}^{2}}\,. 
\end{equation}
Note that the horizon radius $r_h$ is a function of the Hawking temperature, Eq.\eqref{HawkingTemperature}, so that
\begin{equation}
\label{rh_T_Relation}
    r_{h}=\frac{\pi T}{\left|g(x)-\frac{\mu^{2}}{k}c(x)\right|}
\end{equation}
Then the admittance, Eq.\eqref{Admittance1}, can be written as
\begin{equation}\label{admiT}
    \chi(\omega)=\frac{2 i\alpha'e^{-x}}{\omega \pi T^{2}}\left|g(x)-\frac{\mu^{2}}{k}c(x)\right|^{2}\,, 
\end{equation}
where we recall that $x=k/r_h^2$, and then one can see that the admittance is a non-trivial function of the temperature $T$. {\sl This is one of  the main results of this work.} Note that this result is independent of the sign of the chemical potential although it will be relevant in the correlations functions to be calculated in the next section. 

Using the fact that the admittance can be written as \cite{Giataganas:2018ekx}
\begin{equation}
     \chi(\omega)=2\pi\alpha'\left(\frac{i}{\gamma \omega} - \frac{\Delta m}{\gamma^{2}} +\mathcal{O}(\omega)\right),
\end{equation}
we find in our case
\begin{align}
    \gamma&=\frac{e^{-\frac{k}{r_{h}^{2}}}}{r_{h}^{2}}\left(1+\frac{k}{r_{b}^2}+O\left(\frac{k^{2}}{r_{b}^{4}}\right)\right),\\ 
    \Delta m&=\frac{r_{h}^{4}e^{\frac{2k}{r_{h}^{2}}}}{r_{b}^{3}}\left(1+\frac{k}{r_{b}^2}+O\left(\frac{k^{2}}{r_{b}^{4}}\right)\right), 
\end{align}
where $\gamma$ is the friction coefficient and $\Delta m$ corresponds to the change in the bare mass $m$ of the particle described by  the Langevin equation \cite{Kubo1966,Kubo1991}
(see also \cite{Banerjee:2015vmo}). 

If one considers the admittance for case of the string in pure AdS-Schwarzschild, one finds \cite{deBoer:2008gu} 
\begin{equation}\label{AdmittanceAdS}
    \chi_{AdS}(\omega)=\frac{2i\alpha'}{\omega\pi T^{2}}.
\end{equation}

\begin{figure}[ht]
\vskip 0.5cm
	\centering
	\includegraphics[scale = 0.33]{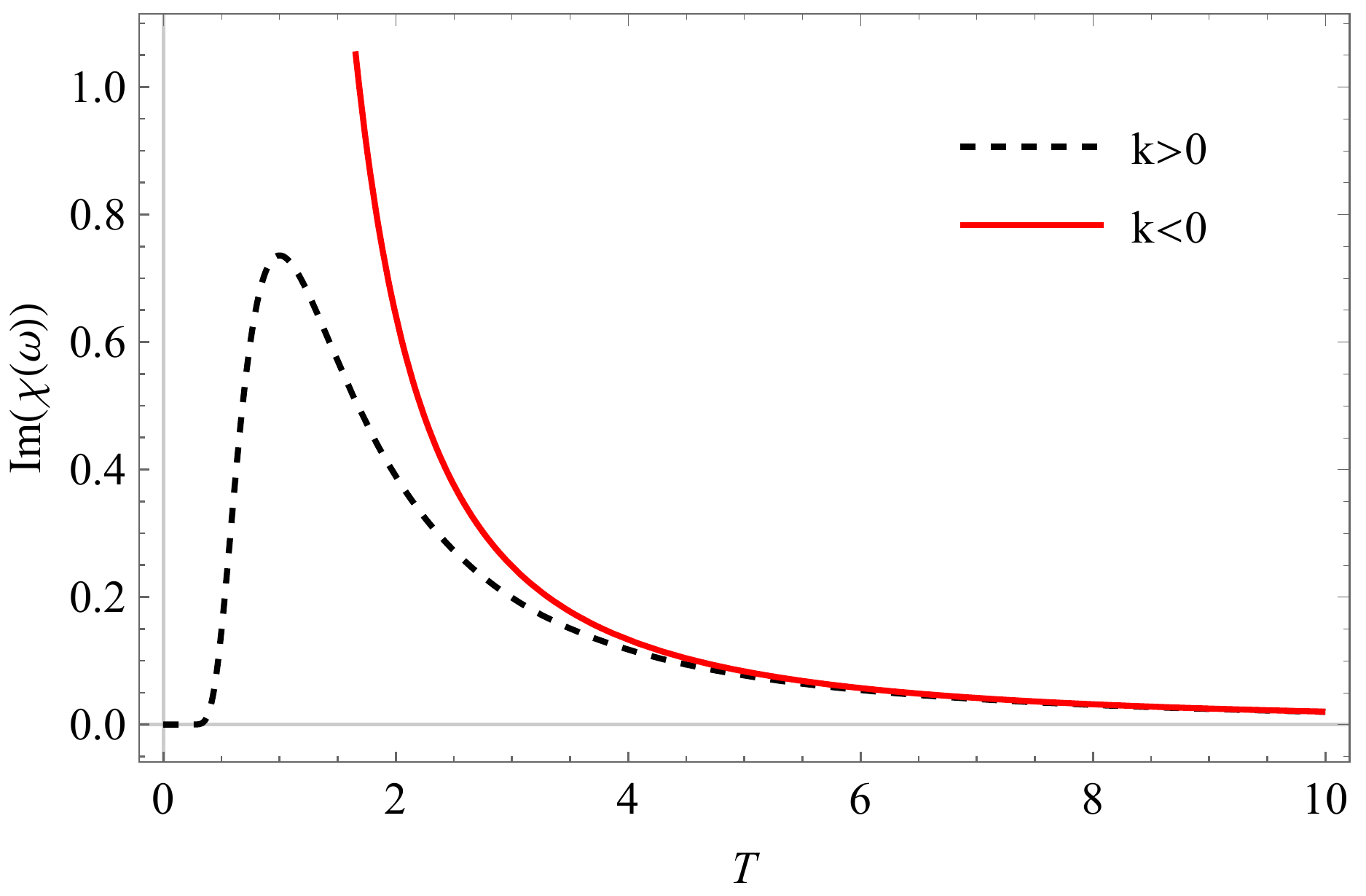}
	\includegraphics[scale = 0.33]{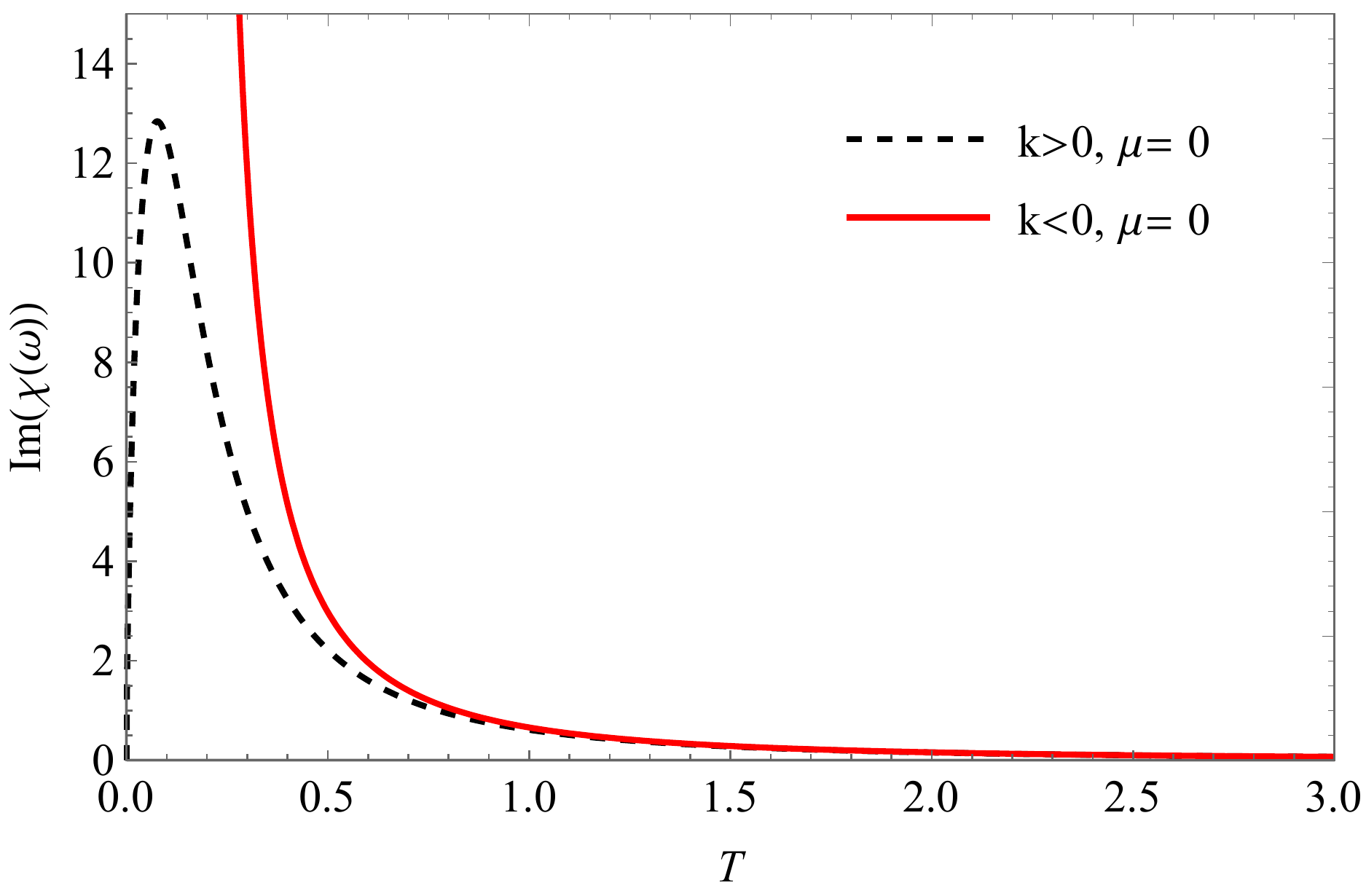}
	\includegraphics[scale = 0.3]{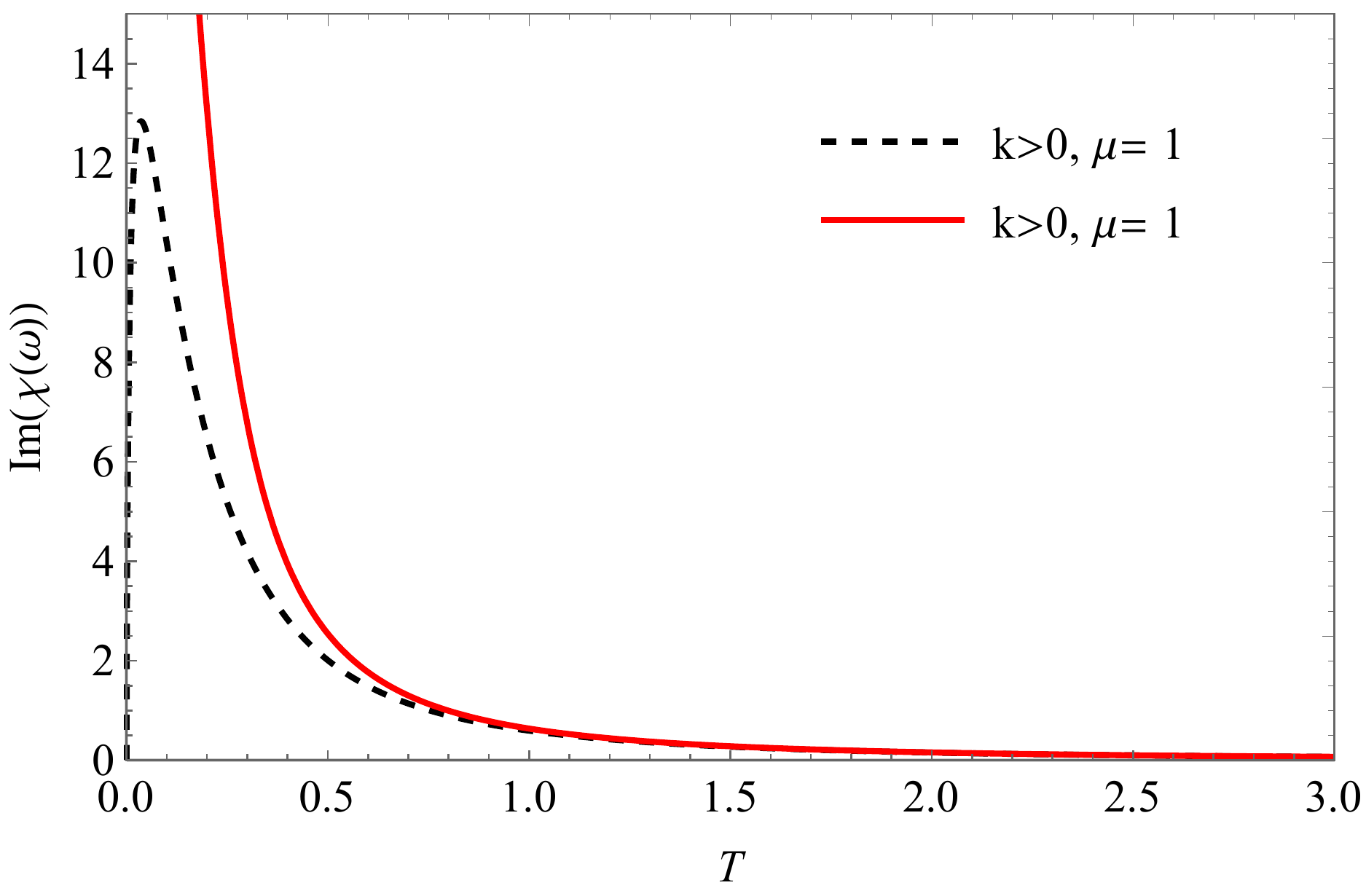}
	\includegraphics[scale = 0.33]{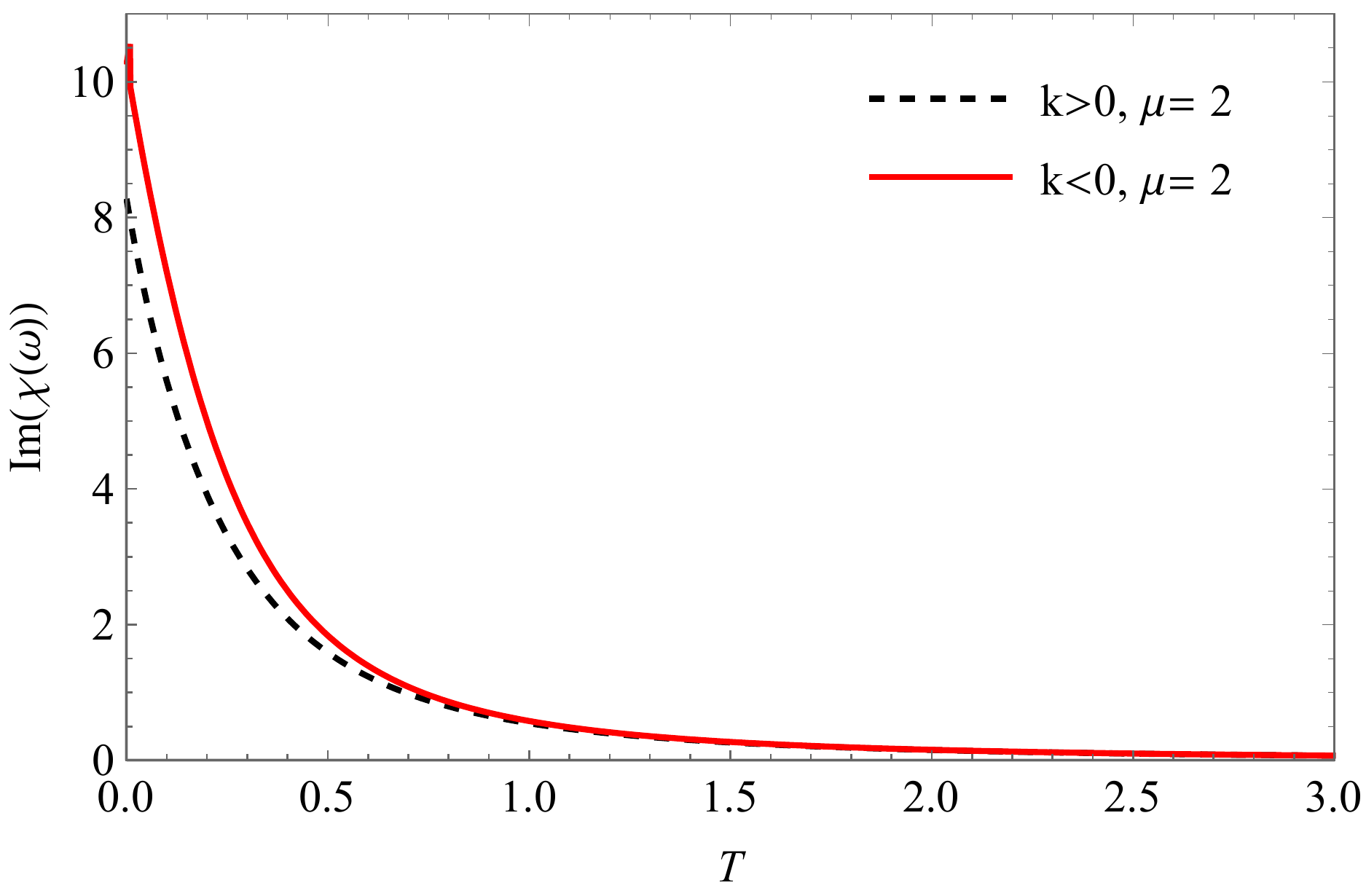}
	\vskip -0.5cm
	\caption{Imaginary part of the admittance,  as a function of the temperature $(T)$ for  $k=\pm 0.18$, in various scenarios. {\sl Upper left panel:} results from Ref. \cite{Caldeira:2020sot} where the backreaction contributions were not considered. {\sl Upper right panel:}  Results  from Ref. \cite{Caldeira:2020rir} where the backreaction was considered although for zero chemical potential. {\sl Lower left and right panels:} Results achieved in this work, Eq. \eqref{admiT}, considering both backreaction and finite chemical potential with $\mu=1$ and $\mu=2$, respectively.}
	\label{many}
\end{figure}

In Refs. \cite{Caldeira:2020sot, Caldeira:2020rir}, the problem of a probe string  immersed in a black hole metric deformed by  an exponential conformal factor $\exp(k/r^2)$ was studied. Regarding backreaction effects, the Ref. \cite{Caldeira:2020sot} did not consider them and the admittance found was: 
\begin{equation}
\label{AdmittanceNoBR} 
     \chi_{NBR}(\omega)=\frac{2i\alpha'e^{-x}}{\omega\pi T^{2}}\,, \qquad {\rm (No\,\, Backreaction)}\,, 
\end{equation}
while in Ref.  \cite{Caldeira:2020rir}, taking into account backreaction, it was obtained that 
\begin{equation}
\label{AdmittanceBR} 
     \chi_{BR}(\omega)=\frac{2i\alpha'e^{-x}}{\omega\pi T^{2}}(g(x))^2\,, \qquad {\rm ( Backreacted)}.
\end{equation}
Note that these results are particular cases of  the admittance $\chi(\omega)$ computed in the present work, Eq. \eqref{admiT}, where finite density and an exponential deformation with  backreaction were included, so that  
\begin{eqnarray}
    \chi(\omega)  &=& \chi_{AdS}(\omega)\left|g(x)-\frac{\mu^{2}}{k}c(x)\right|^{2}e^{-x}\;; \\ \label{Admittance_cases} 
    &=& \chi_{NBR}(\omega)\left|g(x)-\frac{\mu^{2}}{k}c(x)\right|^{2}\;; \label{Admittance_cases2} \\
                    &=& \chi_{BR}(\omega)\left|1-\frac{\mu^{2}}{k}\frac{c(x)}{g(x)}\right|^{2}\;.  \label{Admittance_cases3}
\end{eqnarray}
These equations show the influence of the chemical potential $\mu$ on the admittance $\chi(\omega)$. 
In particular, Eq. \eqref{Admittance_cases3}, displays this extension codified in the term $\mu^2c(x)/k g(x)$. The admittance given by Eqs. \eqref{AdmittanceNoBR}, \eqref{AdmittanceBR}, and 
\eqref{admiT} are depicted in Fig. \ref{many}. This picture shows that the presence of a finite chemical potential and its increase dislocates the curve of the imaginary part of the admittance towards low temperatures.

%


\subsection{Correlation functions and mean square displacement with chemical potential}\label{B}

In this section,  we will calculate the correlation functions and the mean square displacement of the string endpoint on the brane with non-zero chemical potential. 
Starting with the definition of the density operator in the grand canonical ensemble
\begin{equation}
\rho_{0}=\frac{e^{-\beta(\sum_{\omega>0}\omega a_{\omega}^{\dagger}a_{\omega}-\mu a_{\omega}^{\dagger}a_{\omega}})}{\text{Tr}\left( \, e^{-\beta( H-\mu N)}\right)}\,,    
\end{equation}
where the creation and annihilation operators satisfy 
\begin{eqnarray}
\label{ValorEsperado}
\langle a^{\dagger}_{\omega}a_{\omega} \rangle
= \frac{\delta_{\omega\omega'}}{e^{\beta\left(\omega-\mu\right)}\mp 1}; \qquad 
\langle a^{\dagger}_{\omega}a^{\dagger}_{\omega} \rangle
=0; \qquad \langle a_{\omega}a_{\omega} \rangle
=0,
\end{eqnarray}
and the minus (plus) sign corresponds to the bosonic (fermionic) case, one can calculate 
 the correlation function:
\begin{align}
\label{eq:CorrelationPosition}
\langle x(t)x(0) \rangle& \equiv \langle X(t,r_{b})X(0,r_{b}) \rangle\nonumber \\
&=\sum_{\omega>0}\sum_{\omega'>0}
\Big( \frac{h^{*}_{\omega}(r_{b})h_{\omega'}(r_{b}) e^{i\omega t}
+ h_{\omega}(r_{b})h^{*}_{\omega'}(r_{b})e^{-i\omega  t}}{e^{\beta\left(\omega-\mu\right)} \mp 1}
+ h_{\omega}(r_{b})h^{*}_{\omega'}(r_{b})e^{-i\omega  t}\Big)\delta_{\omega\omega'}\nonumber \\
&=\frac{\pi\alpha'r_{h}^{2}f'(r_{h})}{ 2\log\left(\frac{1}{\epsilon}\right)}\sum_{\omega>0}\frac{1}{\omega}
\frac{e^{-\frac{k}{r^{2}_{h}}}}{r^{2}_{h}}\left\{\left(1+\frac{\omega^{2}r^{4}_{h}e^{\frac{2k}{r^{2}_{h}}}}{9r_{b}^{6}}\right)\right.
\cr
& \quad \left. +\, \Re\left[B^{*}\left(1+\frac{\omega^{2}r^{4}_{h}e^{\frac{2k}{r^{2}_{h}}}}{9r_{b}^{6}}-\frac{2i\omega r^{2}_{h}e^{\frac{k}{r^{2}_{h}}}}{3r_{b}^{3}}\right)\right]\right\} 
\left( \frac{2\cos(\omega t)}{e^{\beta\left(\omega-\mu\right)}\mp 1}
+ e^{-i\omega  t}\right)\,,
\end{align}
where we used the solution in terms of the ingoing and outgoing modes $ h_{\omega}(r)=A[h^{out}_{\omega}(r)+Bh^{in}_{\omega}(r)]$ (see the Appendix \ref{apB}). 
Then, we get: 
\begin{align}
\langle x(t)x(0) \rangle&=\frac{\pi\alpha'r_{h}^{2}f'(r_{h})}{\log\left(\frac{1}{\epsilon}\right)}\sum_{\omega>0}\frac{1}{\omega}
\frac{e^{-\frac{k}{r^{2}_{h}}}}{r^{2}_{h}}\left( \frac{2\cos(\omega t)}{e^{\beta\left(\omega-\mu \right)}\mp 1}
+ e^{-i\omega  t}\right)\,.
\end{align}
This sum can be approximated by an integral. For this purpose we use that 
$\Delta\omega={\pi r_{h}^{2}f'(r_{h})}/{\log\left({\epsilon^{-1}}\right)} \sim d\omega \,,$
so that we write 
\begin{eqnarray}
\label{TwoPoint1}
\langle x(t)x(0)\rangle
&=&\frac{\alpha' e^{-\frac{k}{r_{h}^{2}}}}{r^{2}_{h}}
\int_{0}^{\infty}\frac{d\omega}{\omega}
\left( \frac{2\cos(\omega t)}{e^{\beta\left(\omega-\mu\right)}\mp 1}
+ e^{-i\omega  t}\right)
\\ 
\cr 
&=&\langle x(0)x(t)\rangle^*\,.\label{TwoPoint2}
\end{eqnarray}

In a similar way we get that
\begin{align}
\label{eq:CorrelationPositionTime}
\langle x(t)x(t)\rangle
&=\langle X(t,r_{b})X(t,r_{b}) \rangle
\nonumber \\
&= \sum_{\omega>0}
|h^{C}_{\omega}(r_{b})|^{2}
\Big( \frac{2}{e^{\beta(\omega-\mu)}\mp1}
+ 1\Big)
\nonumber \\
&=\frac{\alpha' e^{-\frac{k}{r_{h}^{2}}}}{r^{2}_{h}}\int_{0}^{\infty}\frac{d\omega}{\omega}\left( \frac{2}{e^{\beta(\omega-\mu)}\mp1}
+ 1\right)
\nonumber\\
&=\langle x(0)x(0) \rangle
\end{align}

Now, we can compute the regularized mean square displacement defined by: 
\begin{equation}
    s^2_{\rm reg}(t)
    = \langle
    : [x(t)-x(0)]^2 : \rangle
    \equiv \langle : 
    [X(t, r_b)-X(0, r_b)]^2 : \rangle\,,
\end{equation}{}
where we used the notation $:  [\cdots] :$ to represent the normal ordering prescription.  

Plugging Eqs. \eqref{TwoPoint1}, \eqref{TwoPoint2} and \eqref{eq:CorrelationPositionTime} into the definition of the mean square displacement, we find: 
\begin{equation}
\label{sreggeral}
     s^2_{\rm reg}(t) =\frac{\alpha' e^{-\frac{k}{r_{h}^{2}}}}{r^{2}_{h}}\int_{0}^{\infty}\frac{d\omega}{\omega}\left( \frac{4(1-\cos(\omega t))}{e^{\beta(\omega-\mu)}\mp1}\right)=\frac{8\alpha' e^{-\frac{k}{r_{h}^{2}}}}{r^{2}_{h}}\int_{0}^{\infty}\frac{d\omega}{\omega}\left( \frac{\sin^{2}(\frac{\omega t}{2})}{e^{\beta(\omega-\mu)}\mp 1}\right).
\end{equation}
Up to now, we have been considering the bosonic and fermionic together. From now on, we will split this analysis into two separate cases since the sign of the chemical potential will be crucial.  

\subsubsection{Bosonic case}

In order to calculate the mean square displacement given by Eq. \eqref{sreggeral}, 
for the bosonic case, we take $\mu < 0$, and consider the series expansion 
\begin{equation}\label{eq:geometricseries}
   \frac{1}{e^{\beta( \omega-\mu)}- 1}  =  \frac{e^{-\beta( \omega-\mu)}}{1- e^{-\beta( \omega-\mu)}} = \sum_{n = 0}^{\infty}e^{-\beta (\omega-\mu) (n+1)}\,.
\end{equation}
Then,  the bosonic regularized  mean square displacement reads 
\begin{equation}
s^2_{\rm Breg}(t) =\frac{8\alpha' e^{-\frac{k}{r_{h}^{2}}}}{r^{2}_{h}}\sum_{n=1}^{\infty}\int_{0}^{\infty}\frac{d\omega}{\omega}e^{-\beta (\omega-\mu) n}\sin^{2}(\frac{\omega t}{2}).
\end{equation}
Performing the integral, one gets 
\begin{equation}
\label{S2Series}
s^2_{\rm Breg}(t) =\frac{2\alpha' e^{-\frac{k}{r_{h}^{2}}}}{r^{2}_{h}}\sum_{n=1}^{\infty}e^{\beta\mu n}\log \left(1 + \frac{t^2}{n^2 \beta^2} \right)\,. 
\end{equation}
This expression can be rewritten as  
\begin{eqnarray}
    s^{2}_{\rm Breg}(t)=\frac{2\alpha' e^{-\frac{k}{r_{h}^{2}}}}{r^{2}_{h}}\left(2 {\rm Li_{0}}^{(1,0)}\left(0,e^{\mu }\right) - e^{\mu } \left({\Phi}^{(0,1,0)}\left(e^{\mu },0,1+i \frac{t}{\beta}\right)+{\Phi}^{(0,1,0)}\left(e^{\mu },0,1-i \frac{t}{\beta}\right)\right)\right),
    \cr
\end{eqnarray}
where Li$_n^{(1,0)}(x,y)$ is the first derivative of the polylogarithm function of order $n$ with respect to its first argument $x$ and $\Phi^{(0,1,0)}(x,y,z)$ is the first derivative of the Lerch transcendent function with respect to the second argument $y$.

\begin{figure}
\vskip 0.5cm
	\centering
	\includegraphics[scale = 0.22]{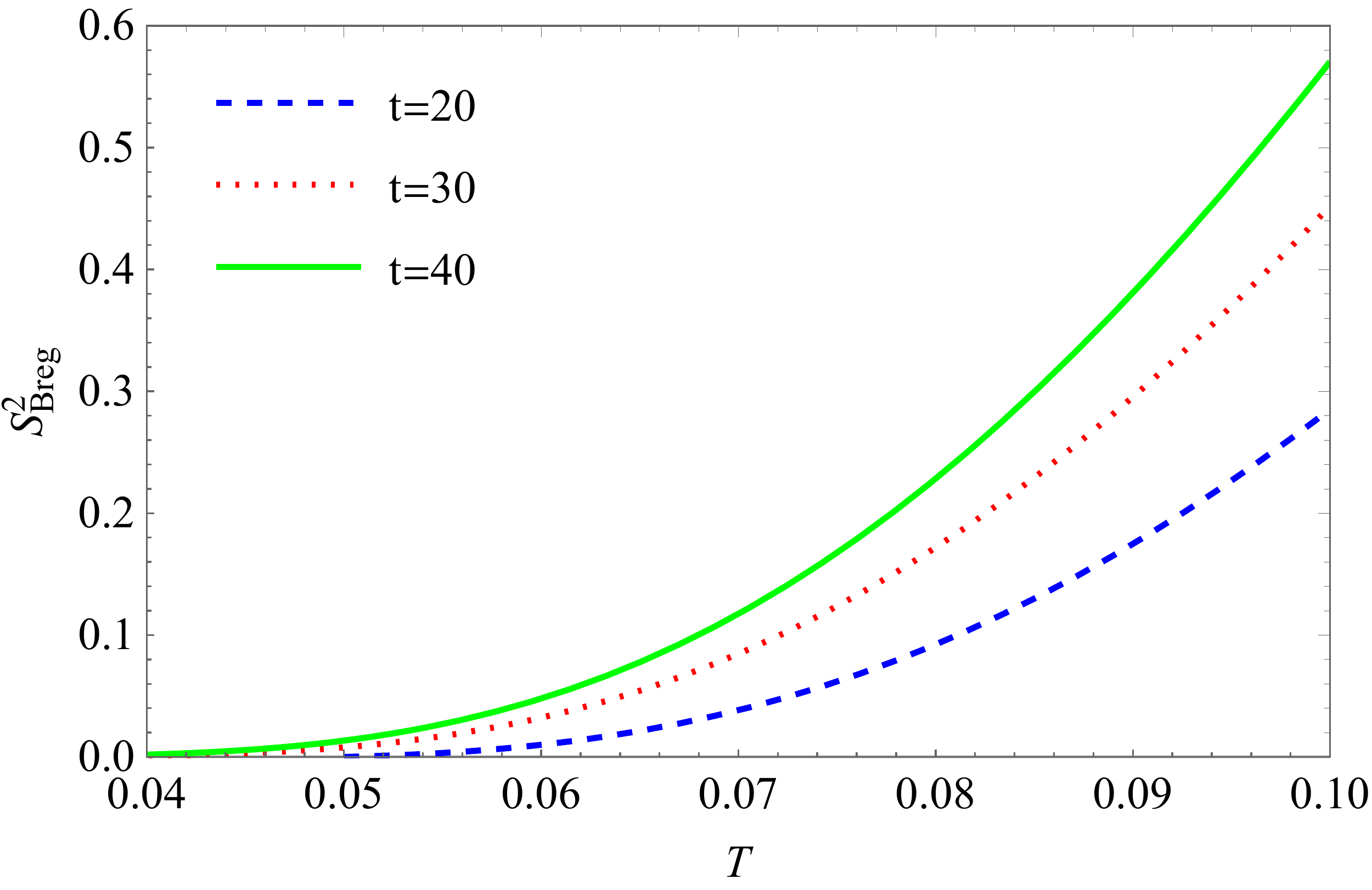}
	\includegraphics[scale = 0.22]{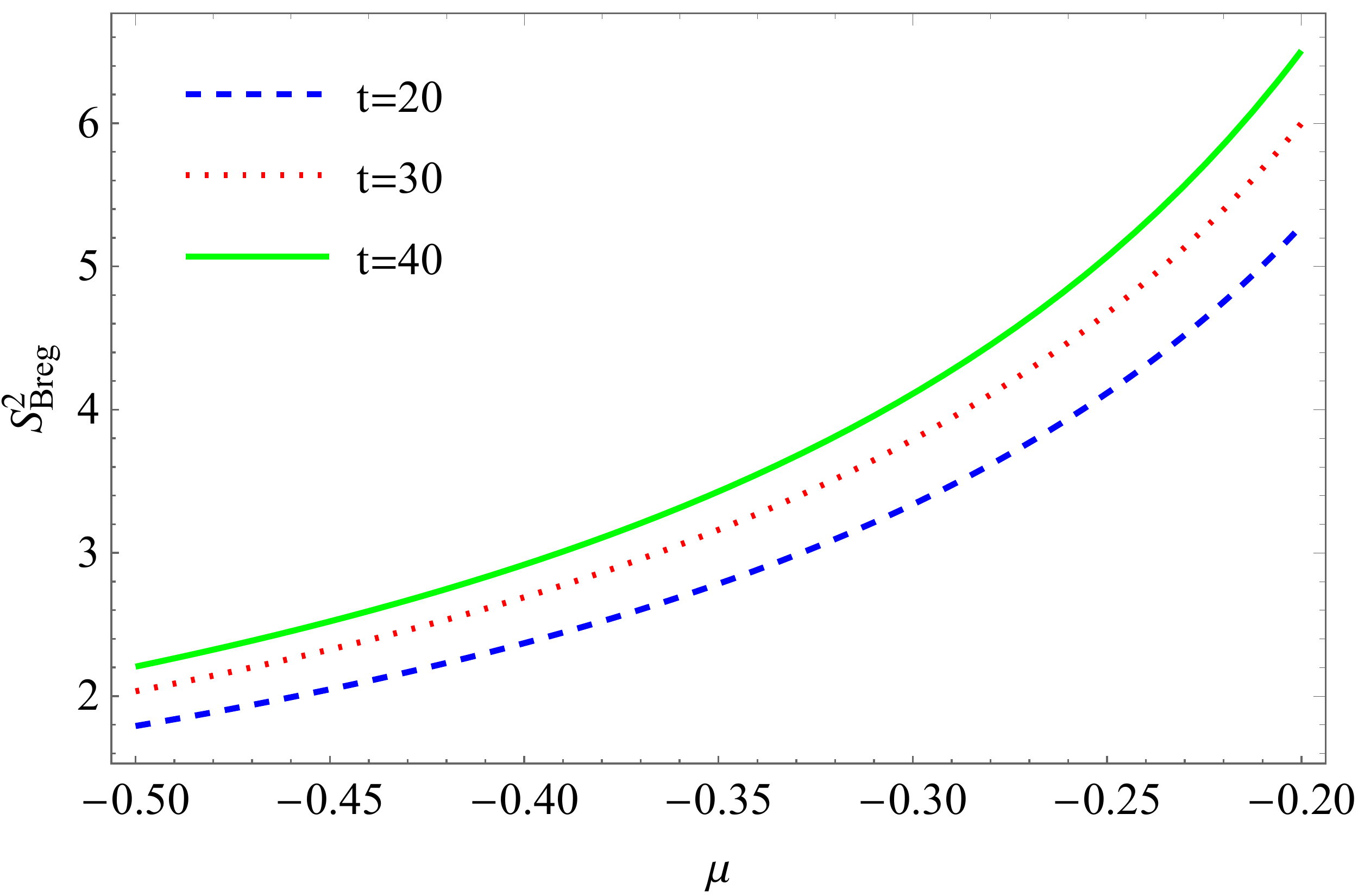}
	\includegraphics[scale = 0.22]{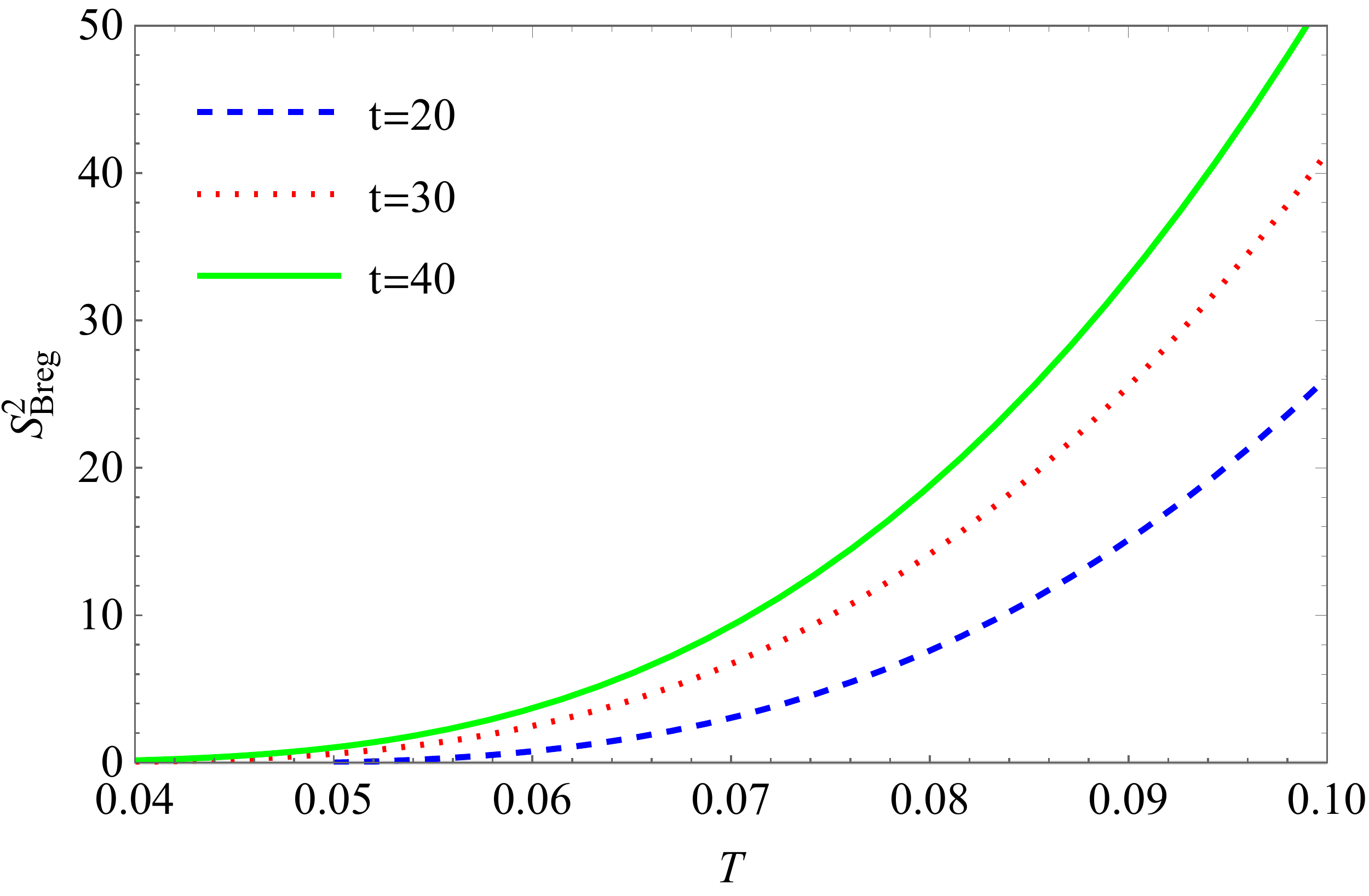}
	\includegraphics[scale = 0.23]{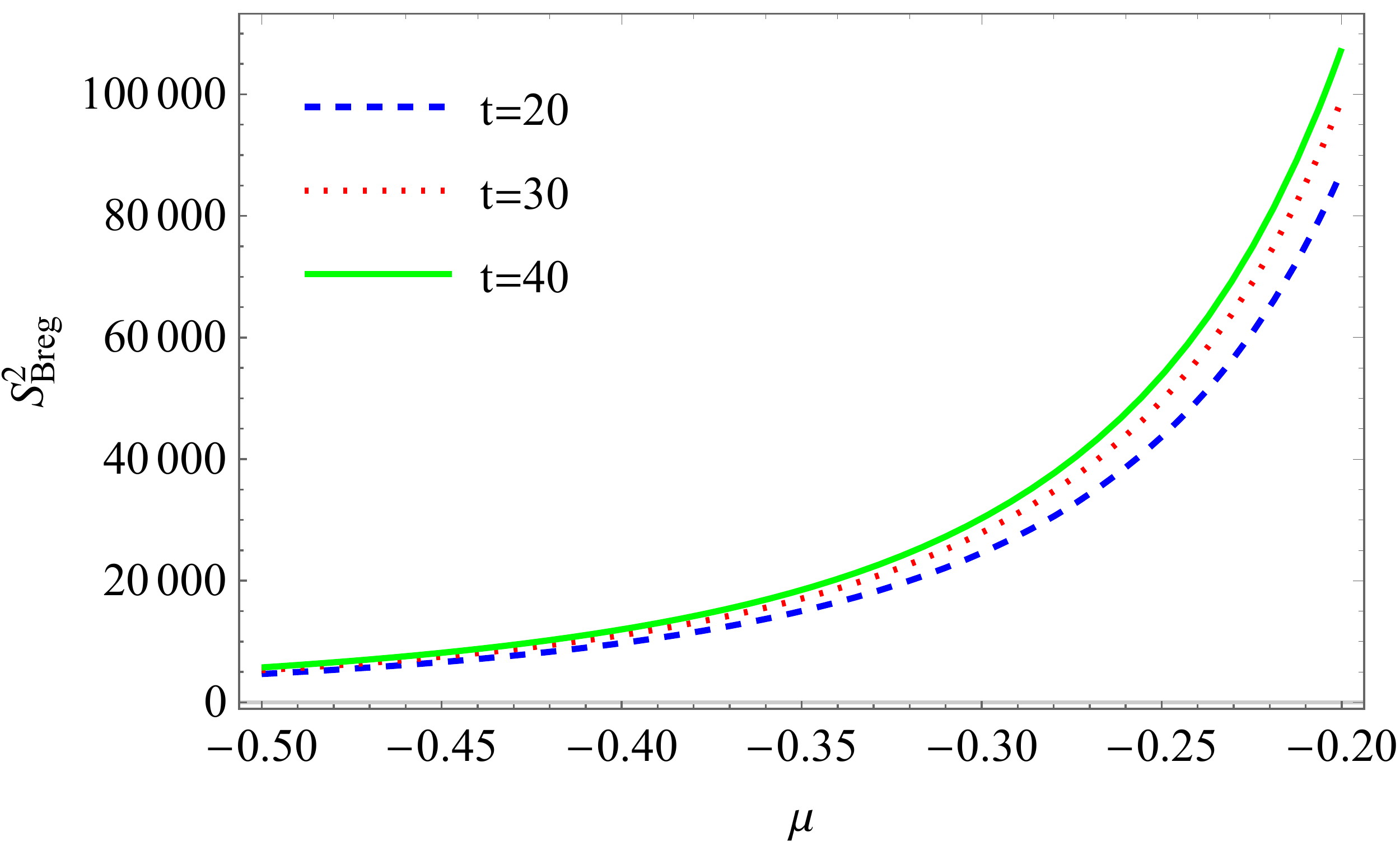}
	\vskip -0.5cm
	\caption{Bosonic regularized mean square displacement $s^2_{\rm Breg}$ as a function of the temperature $T$ or the chemical potential $\mu$ for the late time approximation, Eq. \eqref{bostgran}, for some values of the time $t$. {\sl Upper panels:} positive fixed value of $k$. {\sl Lower panels:} negative fixed value of $k$.}
	\label{figbostgran}
\end{figure}

Now, it is interesting to consider the late time approximation $t\gg\beta$ for Eq. \eqref{S2Series} as 
\begin{eqnarray}
\label{bostgran}
     s^{2}_{\rm Breg}(t)
     &\approx&\frac{2\alpha' e^{-\frac{k}{r_{h}^{2}}}}{r^{2}_{h}}
     \sum_{n=1}^{\infty}e^{\beta\mu n}
     \log \left( \frac{t^2}{n^2 \beta^2} \right)\cr
     &=&\frac{2\alpha' e^{-\frac{k}{r_{h}^{2}}}}{r^{2}_{h}}\left(-\sum_{n=1}^{\infty}e^{\beta\mu n}\log \left( n^2 \right)+\sum_{n=1}^{\infty}e^{\beta\mu n}\log \left( \frac{t^2}{ \beta^2} \right)\right)\cr
    &=&\frac{2\alpha' e^{-\frac{k}{r_{h}^{2}}}}{r^{2}_{h}}\left(-\sum_{n=1}^{\infty}e^{\beta\mu n}\log \left( n^2 \right)+\frac{2}{e^{-\beta\mu}-1}\log \left( \frac{t}{ \beta} \right)\right)\,,
\end{eqnarray}
which converges since $\mu<0$. This result  corresponds to a sub-diffusive regime as $s^2\sim \log t$, as seen in Ref. \cite{sinai-like}.
This behavior is shown in Fig. \ref{figbostgran}, for various values of the time $t$ against the temperature $T$ or the chemical potential $\mu$ for positive and negative $k$. 


 If one considers the limit $\mu/T \to 0$ in Eq. \eqref{S2Series}, one can rewrite that sum as
\begin{eqnarray}
s^2_{\rm Breg}(t)
&=&
\frac{2\alpha' e^{-\frac{k}{r_{h}^{2}}}}{r^{2}_{h}}\sum_{n=1}^{\infty}
e^{\beta\mu n}\log \left(1 + \frac{t^2}{n^2 \beta^2} \right)\,
\cr
&\approx&
\frac{2\alpha' e^{-\frac{k}{r_{h}^{2}}}}{r^{2}_{h}}\sum_{n=1}^{\infty} \log \left(1 + \frac{t^2}{n^2 \beta^2} \right)\,
\cr
&=&
\frac{2\alpha' e^{-\frac{k}{r_{h}^{2}}}}{r^{2}_{h}} \log \left(\frac{\beta\sinh\left(\frac{\pi t}{ \beta}\right)}{\pi t}\right)\,. 
\end{eqnarray}
In this case, for long times, $t \gg \beta/\pi$, one has
\begin{eqnarray}
s^{2}
&=&
\frac{2\alpha' e^{-\frac{k}{r_{h}^{2}}}}{r^{2}_{h}} 
\left[
\log \left(e^{\frac{\pi t}{ \beta}}
-e^{-\frac{\pi t}{ \beta}}\right)
-
\log\left(\frac{2\pi t}{\beta}\right
)\right]
\,
\cr
&\approx&
\frac{2\pi\alpha' e^{-\frac{k}{r_{h}^{2}}}}{\beta r^{2}_{h}}t\,. 
\end{eqnarray}
So, in the limit $\mu/T \to 0$, one recovers the usual linear behavior for  the mean square displacement.

%

\begin{figure}
\vskip 0.5cm
	\centering
	\includegraphics[scale = 0.22]{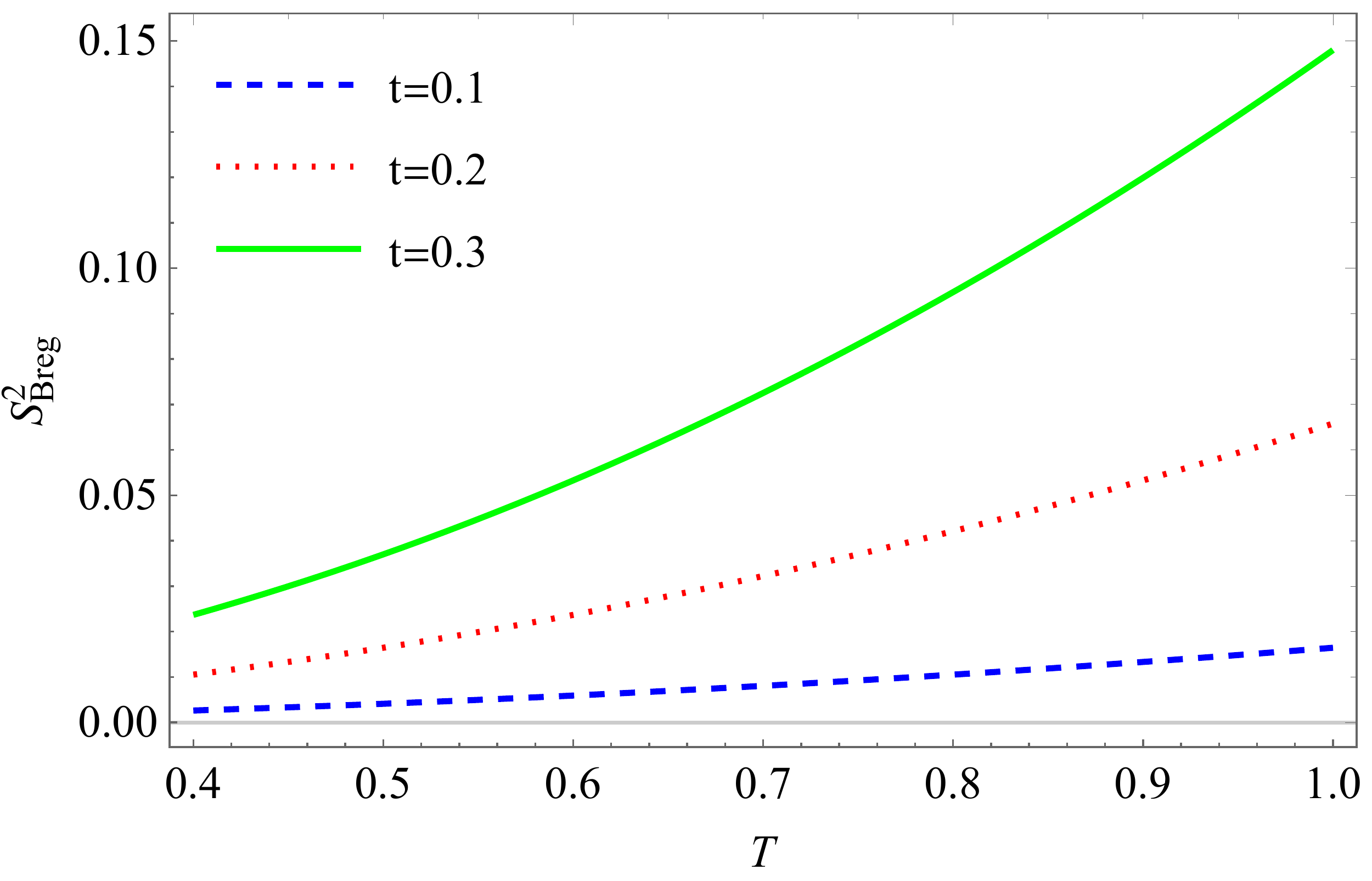}
	\includegraphics[scale = 0.22]{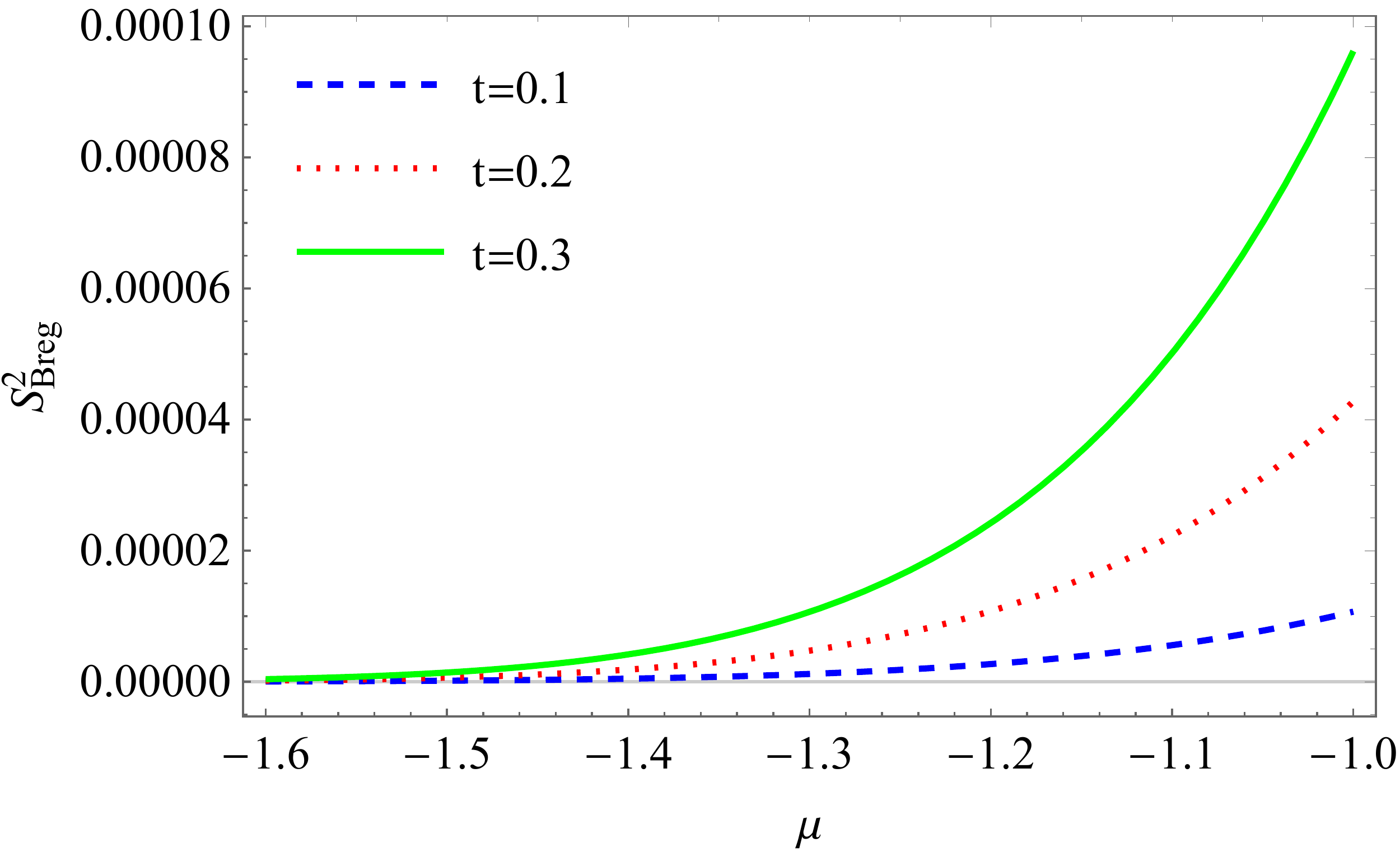}
	\includegraphics[scale = 0.22]{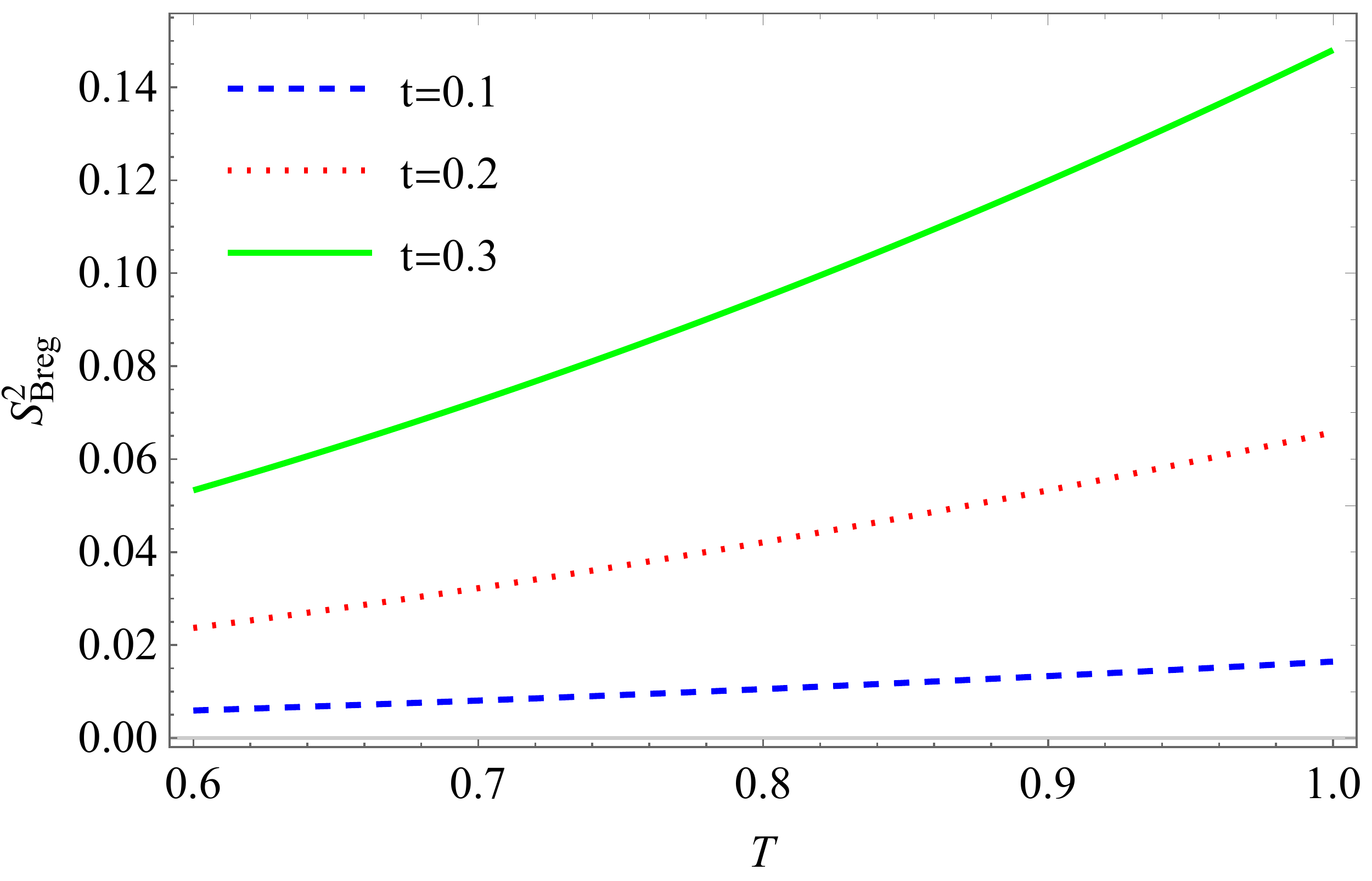}
	\includegraphics[scale = 0.22]{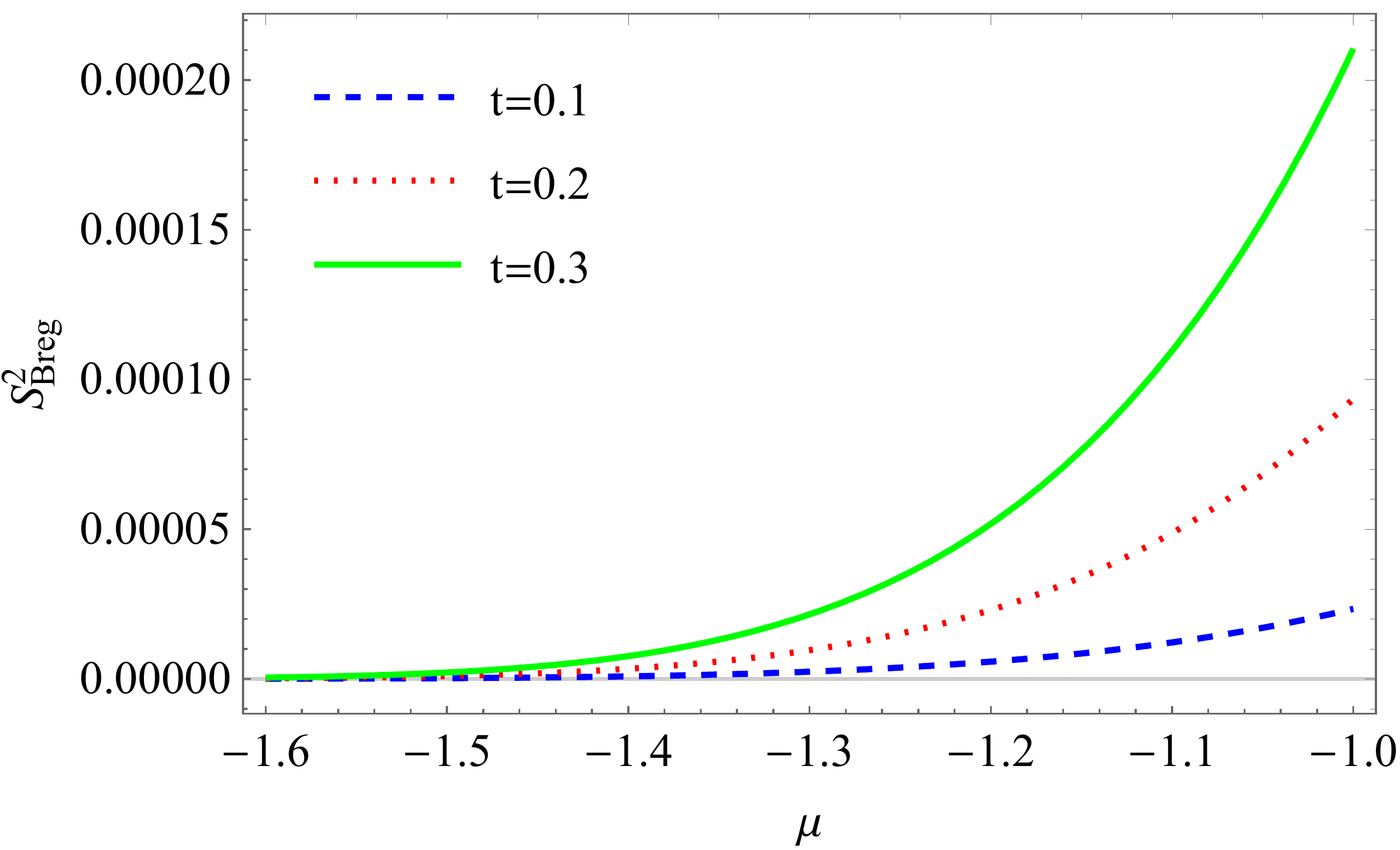}
	\vskip -0.5cm
	\caption{Bosonic regularized mean square displacement $s^2_{\rm Breg}$ as a function of the temperature $T$ or the chemical potential $\mu$ for the short time approximation, Eq. \eqref{bostpeq}, for some values of the time~$t$. {\sl Upper panels:} positive fixed value of $k$. {\sl Lower panels:} negative fixed value of $k$.}
	\label{figbostpeq}
\end{figure}

On the other hand, for the short time approximation $t\ll\beta$, we find
\begin{eqnarray}\label{bostpeq}
     s^{2}_{\rm Breg}(t)
     \approx\frac{t^{2}}{\beta^{2}}
     \left(\sum _{n=1}^{\infty } \frac{e^{\beta\mu  n}}{n^2}\right)= \frac{t^{2}}{\beta^{2}}\, 
     \text{Li}_2\left(e^{\beta\mu }
     \right)\,, 
\end{eqnarray}
where $\text{Li}_2\left(e^{\beta\mu }\right)$ is the  polylogarithm function of order 2. As expected, this corresponds to the ballistic regime which goes like $s^2 \sim t^2$, as  shown in Fig.~\ref{figbostpeq}.
%

\subsubsection{Fermionic case}

Here, we are going to consider the fermionic regularized mean square displacement. From Eq. \eqref{sreggeral}, taking the Fermi Dirac distribution with $\mu > 0$, we have 
\begin{equation}\label{ferm}
     s^2_{\rm Freg}(t)=\frac{8\alpha' e^{-\frac{k}{r_{h}^{2}}}}{r^{2}_{h}}\int_{0}^{\infty}\frac{d\omega}{\omega}\left( \frac{\sin^{2}(\frac{\omega t}{2})}{e^{\beta(\omega-\mu)}+1}\right)\,. 
\end{equation} 
We can calculate this integral using the Sommerfeld expansion:  
\begin{eqnarray}\label{expsome}
    \int_{0}^{\infty}\frac{d\omega}{\omega}\left( \frac{\sin^{2}(\frac{\omega t}{2})}{e^{\beta(\omega-\mu)}+1}\right)&=&\int_{0}^{\mu}\frac{\sin^{2}(\frac{\omega t}{2})}{\omega}d\omega+\left.\frac{\pi^{2}}{6\beta^{2}}\frac{d}{d\omega}
    \left(\frac{\sin^{2}(\frac{\omega t}{2})}{\omega}\right)\right|_{\omega=\mu} \cr
    &=&\frac{1}{2} (-\text{Ci}(t \mu )+\log (t\mu )+\gamma )+\frac{\pi^{2}}{6\beta^{2}}\frac{\mu  t \sin (\mu  t)-2 \sin ^2\left(\frac{\mu  t}{2}\right)}{2 \mu ^2}, 
\end{eqnarray}
where we kept terms up to order ${1}/{\mu^{2}\beta^{2}}$,  $\text{Ci} (z)=-\int_z^{\infty } \frac{ \cos(s) }{s} \, ds$ is the Cosine integral function and $\gamma$  is Euler's constant, with numerical value $\gamma\approx 0.577216$ .

\begin{figure}
\vskip 0.5cm
	\centering
	\includegraphics[scale = 0.22]{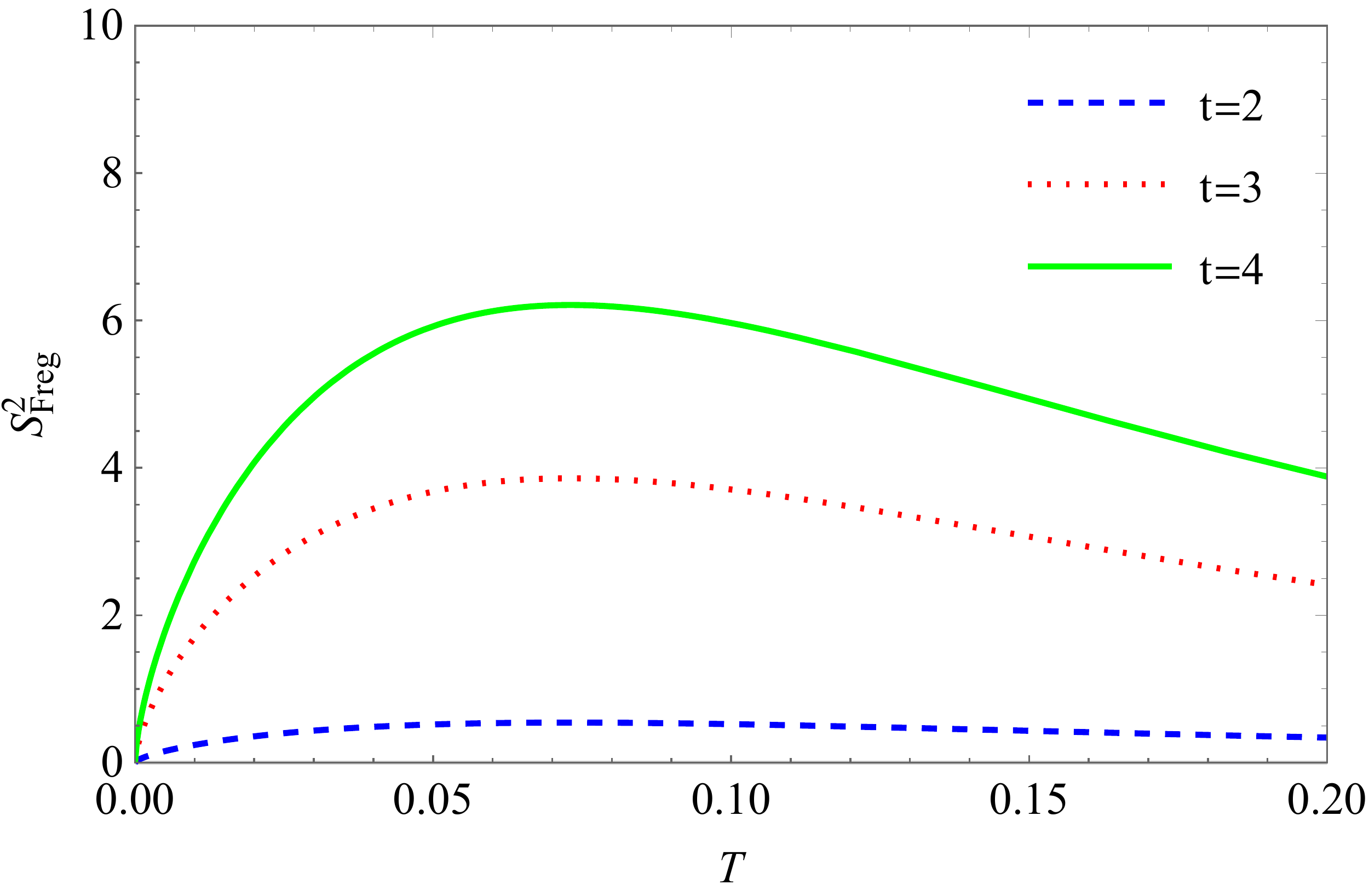}
	\includegraphics[scale = 0.22]{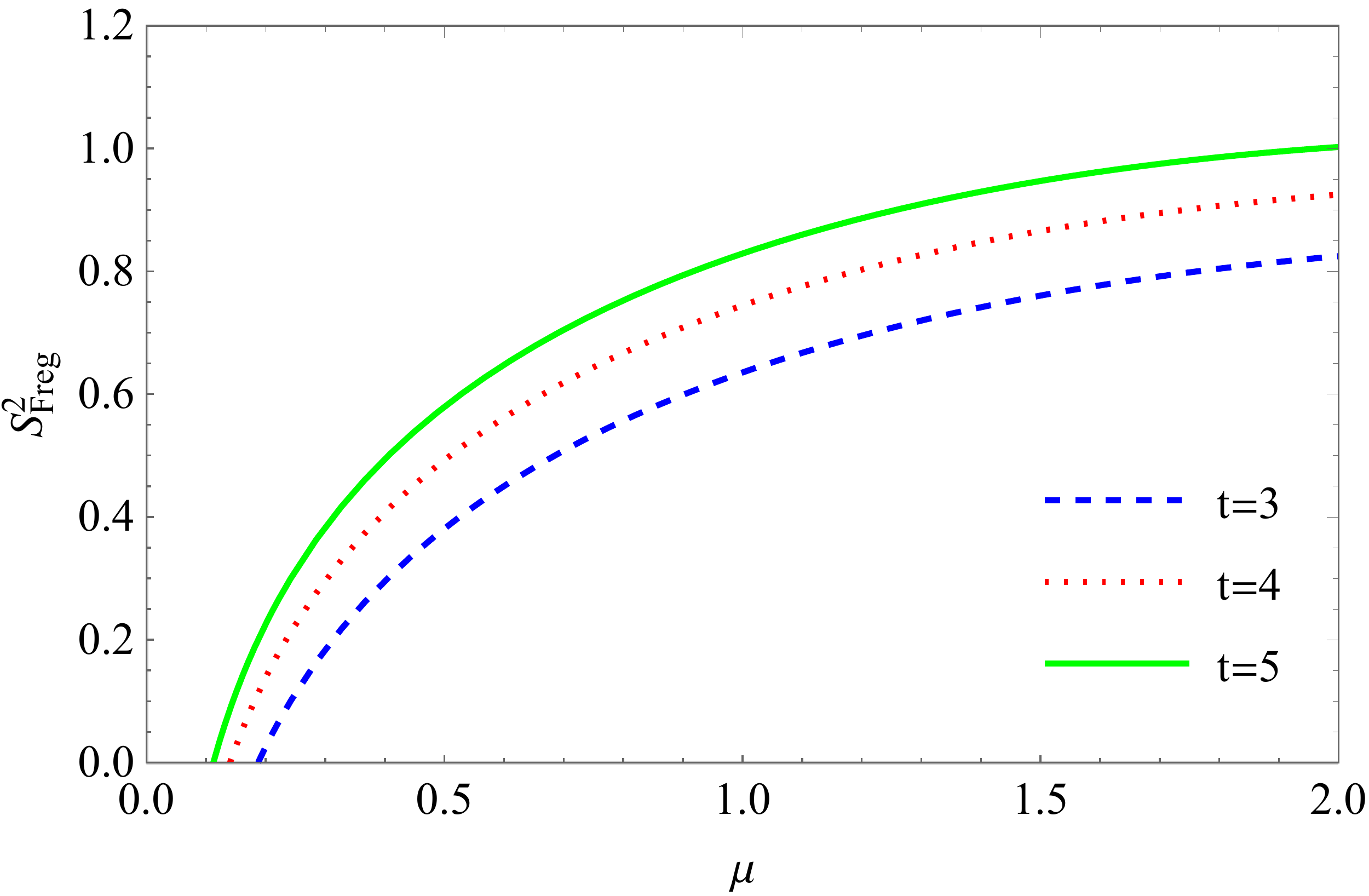}
	\includegraphics[scale = 0.22]{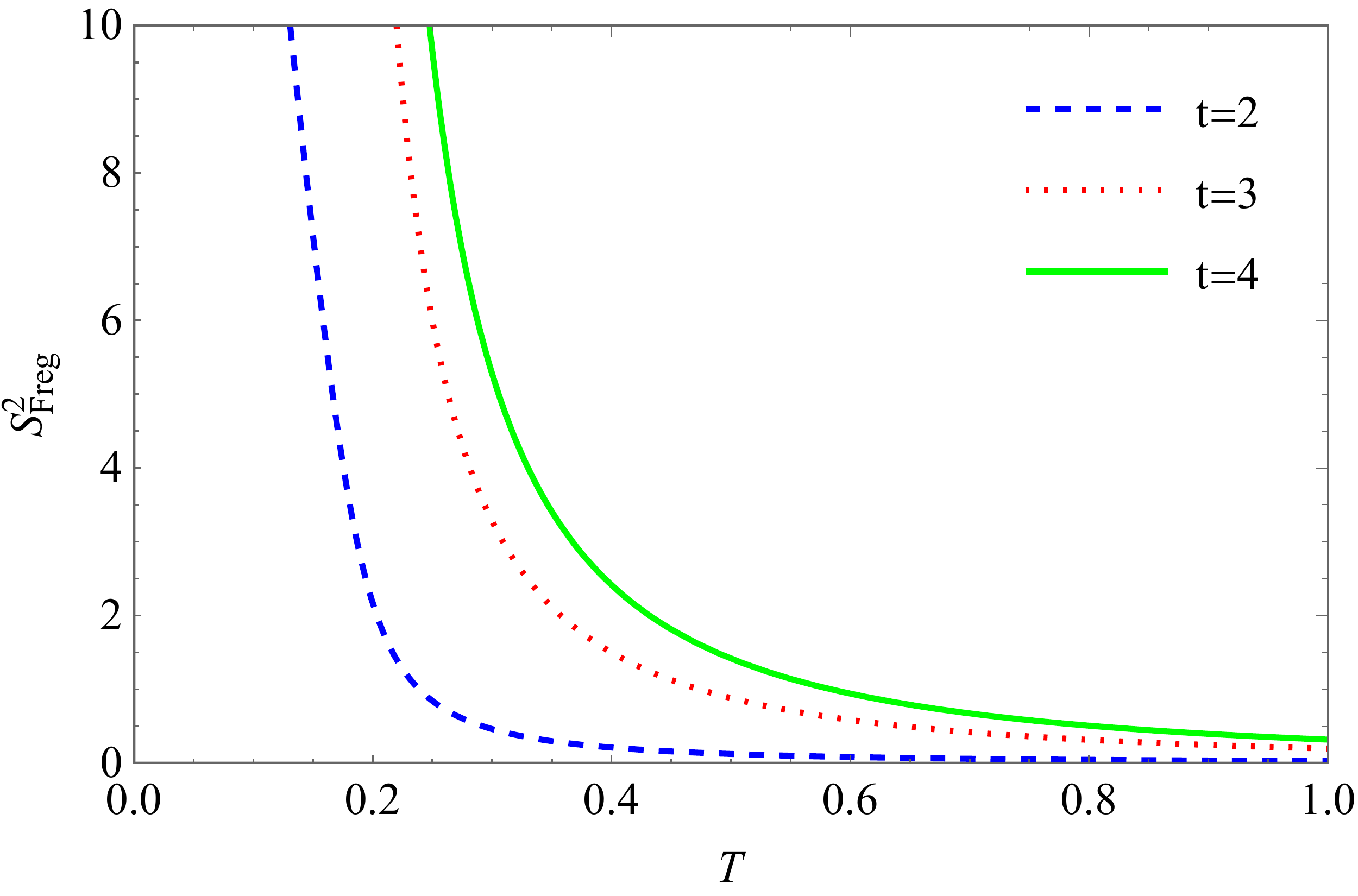}
	\includegraphics[scale = 0.22]{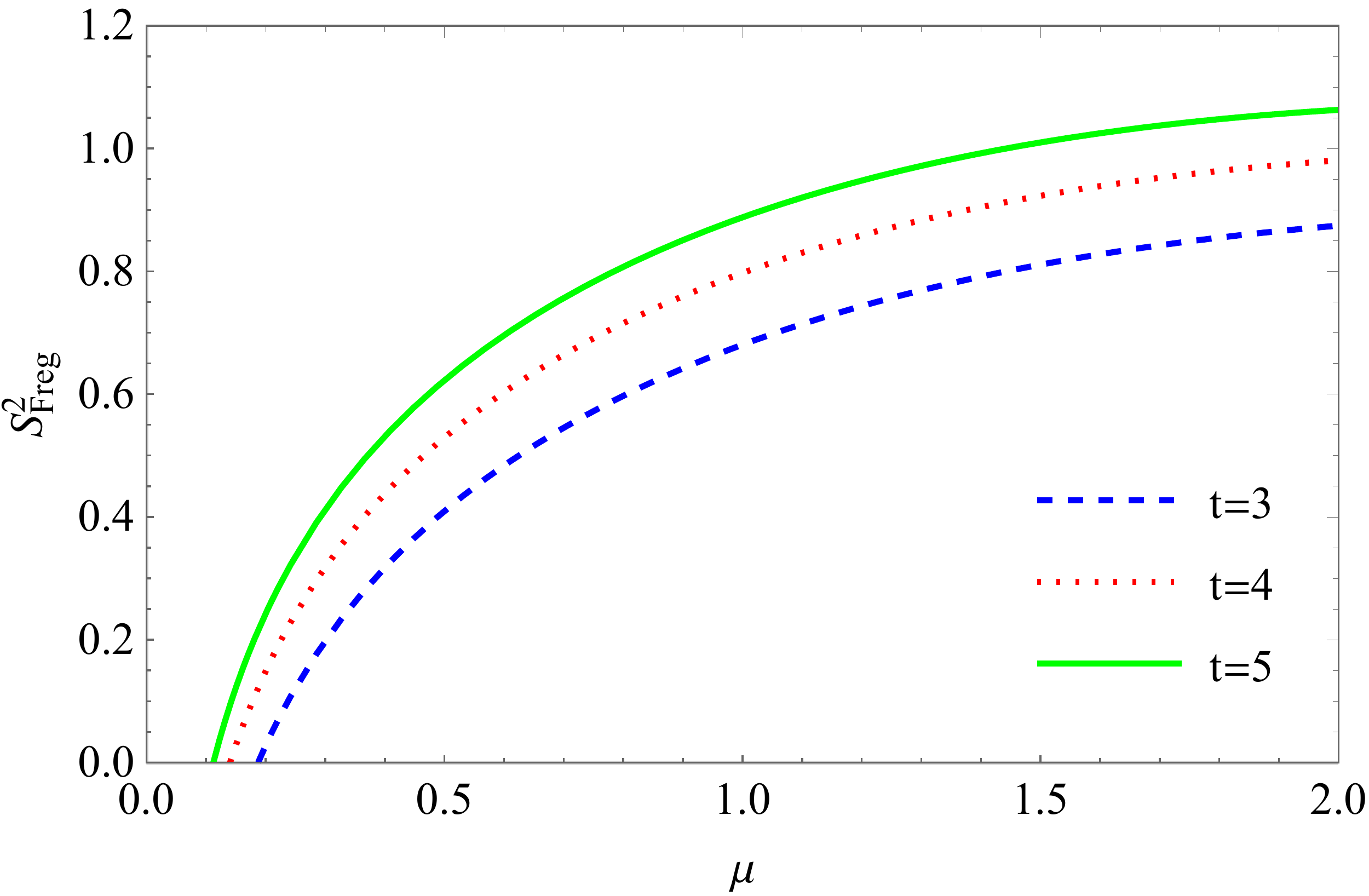}
	\vskip -0.5cm
	\caption{Fermionic regularized mean square displacement $s^2_{\rm Freg}$ as a function of the temperature $T$ or the chemical potential $\mu$ for the large time approximation, Eq. \eqref{fertgran}, for some values of the time~$t$. {\sl Upper panels:} positive fixed value of $k$. {\sl Lower panels:} negative fixed value of $k$.}
	\label{figfertgran}
\end{figure}

In the low temperature regime,  $\mu\beta=\frac{\mu}{T}\gg1$, we can disregard the second term in the above equation and obtain for large times $t \gg \mu$
\begin{equation}\label{fertgran}
     s^2_{\rm Freg}(t) \approx\frac{4\alpha' e^{-\frac{k}{r_{h}^{2}}}}{r^{2}_{h}} \left(\log \left(t\mu \right)+ \gamma \right)
\end{equation}
which is again a sub-diffusive regime
$s^2 \sim \log t$, as in the bosonic case, or the ones seen in Ref. \cite{sinai-like}.
This behavior is shown in Fig. \ref{figfertgran}. 

In order to take the limit $\mu/T \to 0$, here in the fermionic case, one needs to go back to Eq. \eqref{ferm}, before the Sommerfeld expansion Eq. \eqref{expsome}. Since this limit corresponds to a high temperature limit, one sees that the Fermi distribution goes over the classical one, where bosons and fermions are degenerate. In this case the the mean square displacement goes as $s^2\sim t$, as expected \cite{Caldeira:2020sot, Caldeira:2020rir}.

\begin{figure}
\vskip 0.5cm
	\centering
	\includegraphics[scale = 0.22]{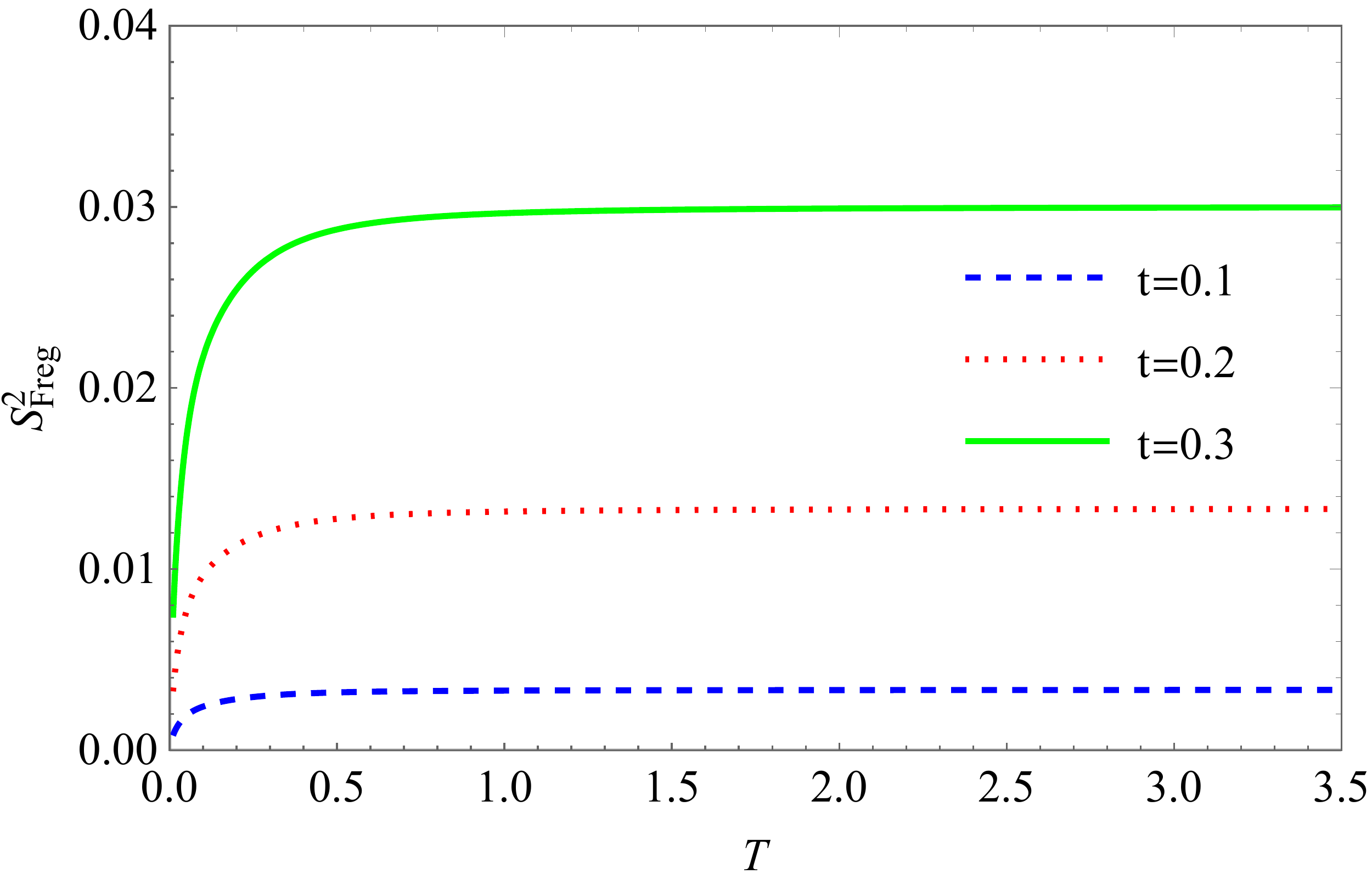}
	\includegraphics[scale = 0.22]{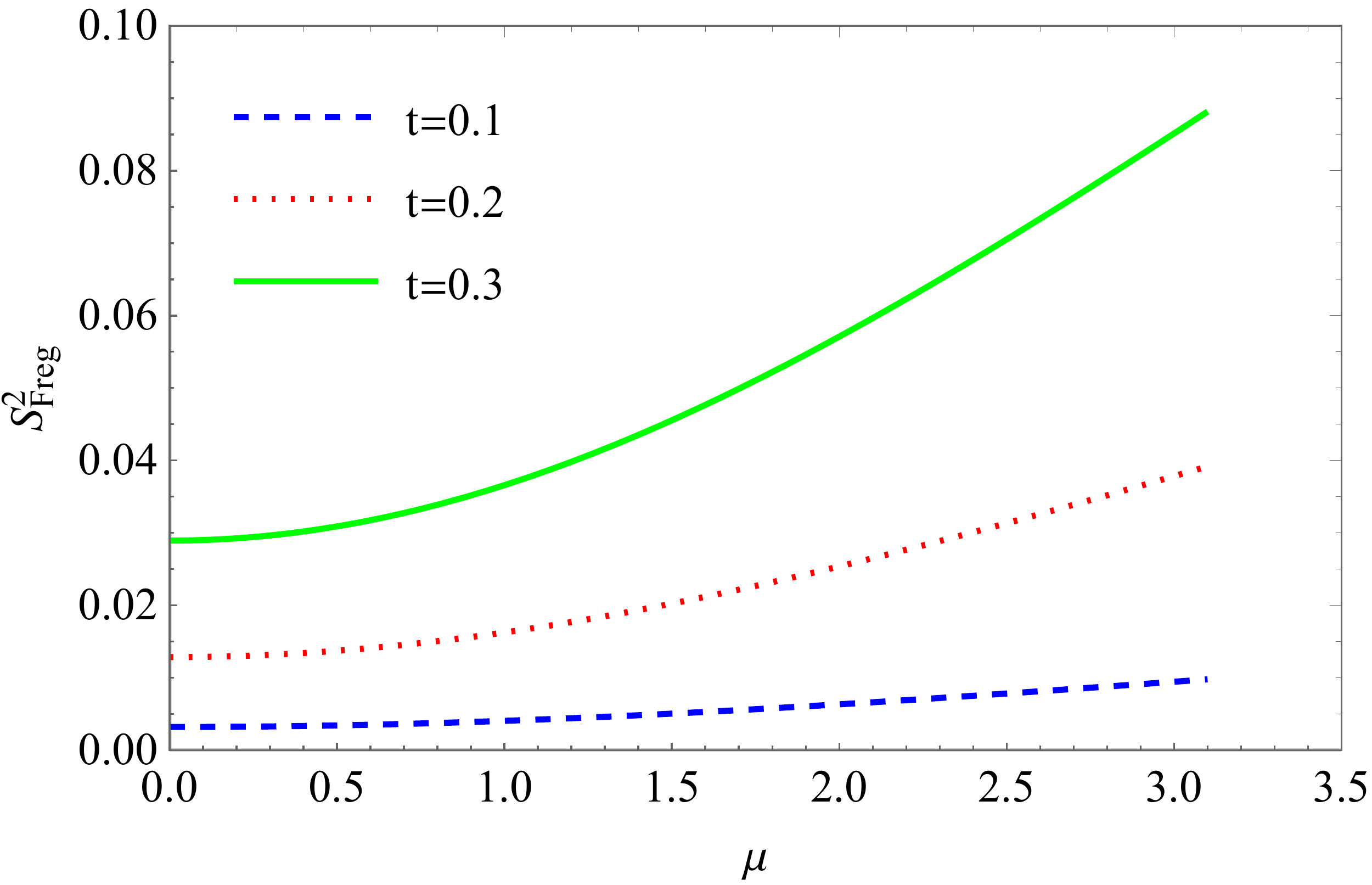}
	\includegraphics[scale = 0.22]{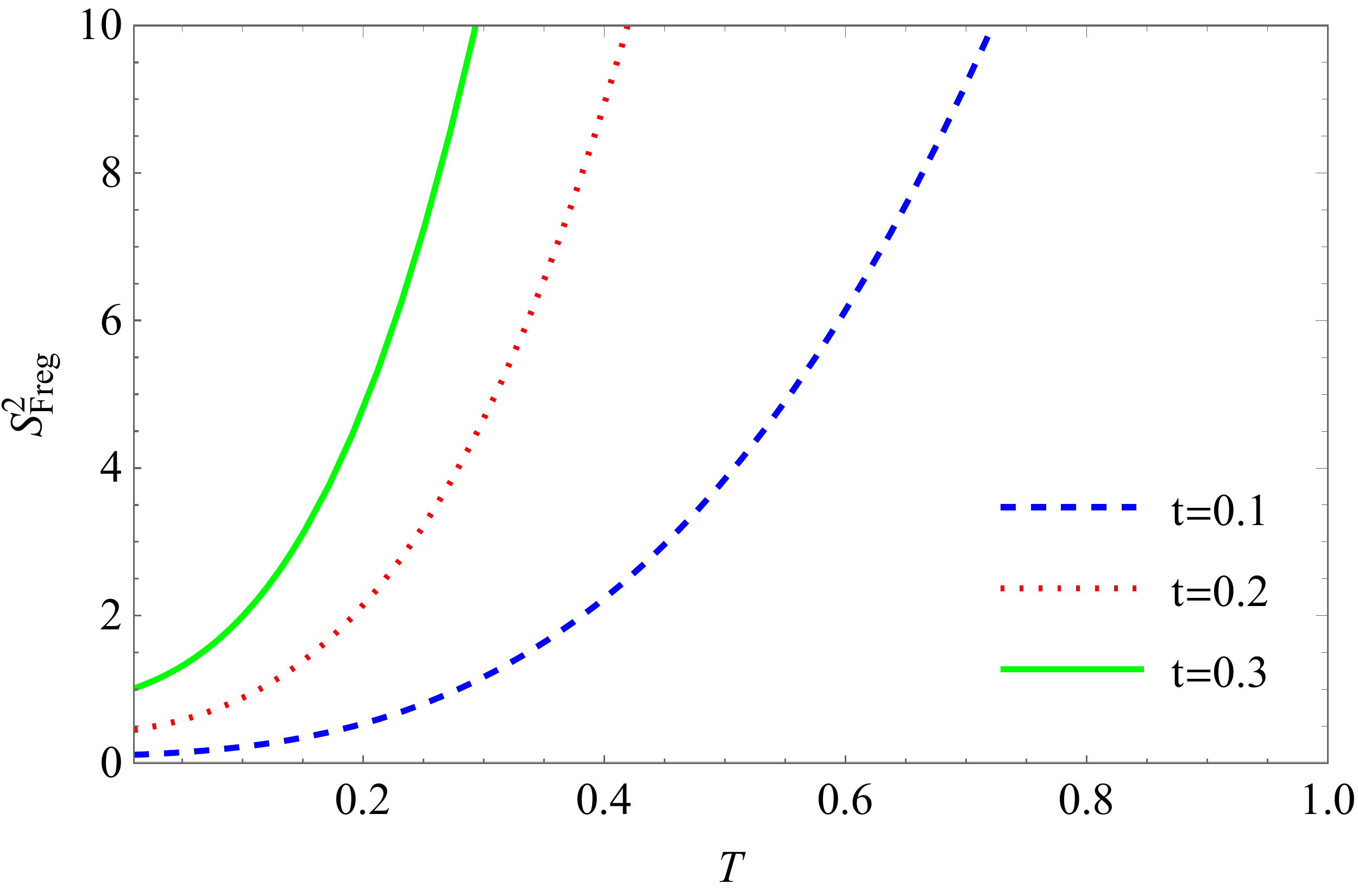}
	\includegraphics[scale = 0.22]{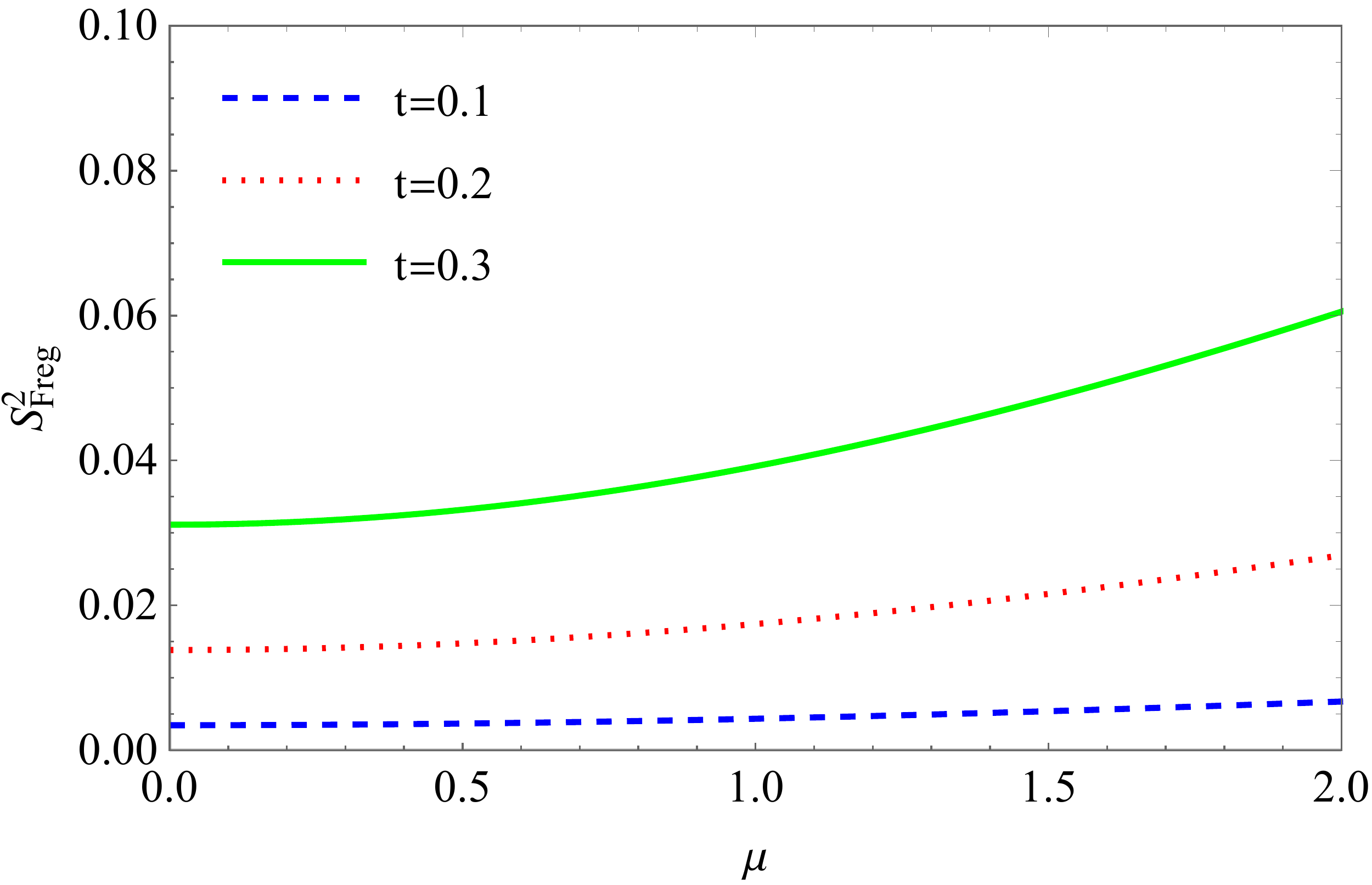}
	\vskip -0.5cm
	\caption{Fermionic regularized mean square displacement $s^2_{\rm Freg}$ as a function of the temperature $T$ or the chemical potential $\mu$ for the short time approximation, Eq. \eqref{fertpeq}, for some values of the time~$t$. {\sl Upper panels:} positive fixed value of $k$. {\sl Lower panels:} negative fixed value of $k$.}
	\label{figfertpeq}
\end{figure}

On the other hand, for small times, $t \ll \mu$, the mean square displacement can be approximated as
\begin{equation}\label{fertpeq}
    s^2_{\rm Freg}(t) \approx\frac{\alpha' e^{-\frac{k}{r_{h}^{2}}}}{r^{2}_{h}}\left(\frac{\pi ^2}{3 \beta ^2}+\mu^{2}\right)t^2\,, 
\end{equation}
which corresponds to a usual ballistic regime $s^2 \sim t^2$. 
From Eq. \eqref{fertpeq} one can see its behavior in Fig. \ref{figfertpeq}. 
 
It is interesting to note that for the   bosonic and fermionic cases,  Eqs. \eqref{bostgran} and \eqref{fertgran},  both have as a large time approximation a sub-diffusive behavior with $s^{2}\sim \log(t)$. 
Such a logarithm time dependence was found  in ultraslow scaled Brownian motion representing physical systems in Sinai diffusion. For instance, in quenched random energy landscapes, random walks on bundled structures, colloidal hard sphere systems at the liquid-glass transition, or granular gases in the homogeneous cooling stage, 
as discussed in Ref. \cite{sinai-like}. 

This is a slower process than  the usual $s^{2}\sim t$, found for the Brownian motion that follows the Langevin equation 
\begin{equation}
    m\ddot{x}
    =-\gamma m\dot{x}
    +R(t)\,, 
\end{equation}
where $\gamma$ is the drag coefficient, $m$ is the mass and $R(t)$ is an stochastic variable representing the random force acting on the particle.

\section{Checking the fluctuation-dissipation theorem for bosonic and fermionic cases}
\label{sec5}

Now, after calculating the admittance and the correlation functions for the bosonic and fermionic systems, we are ready to check the fluctuation-dissipation theorem in both cases, as stated in section \ref{sec2}. 

First of all, we can rewrite the r.h.s of Eqs. \eqref{FDTB} and \eqref{FDTF} as 
\begin{eqnarray} 
\mathcal{F}^{-1}\left[\left(1+2n_{B}\right)\Im \chi (\omega)\right]   &=& \frac{1}{2\pi}\int_{-\infty}^{\infty}d\omega\left(1+\frac{2}{e^{\beta(|\omega|-\mu)}-  1}\right)\Im \chi (|\omega|)e^{i\omega t}\,, \\
 \mathcal{F}^{-1}\left[\left(1+2n_{F}\right)\Im \chi (\omega)\right]   &=& \frac{1}{2\pi}\int_{-\infty}^{\infty}d\omega\left(1+\frac{2}{e^{\beta(|\omega|-\mu)}+ 1}\right)\Im \chi (|\omega|)e^{i\omega t} \,, 
\end{eqnarray}
where we are considering the frequencies as positive in the physical quantities. From the admittance, Eq. \eqref{Admittance1}, the above equations can be calculated as 
\begin{eqnarray}
\label{Vascao}
   \frac{1}{2\pi}\int_{-\infty}^{\infty}d\omega\left(1+\frac{2}{e^{\beta(|\omega|-\mu)}\mp 1}\right)\Im \chi (|\omega|)
   e^{i\omega t} 
 =\frac{\alpha'e^{-\frac{k}{r_{h}^{2}}}
     }{ r_{h}^{2}}\int_{-\infty}^{\infty}\frac{d\omega}{|\omega|}\left(1+\frac{2}{e^{\beta(|\omega|-\mu)}\mp 1}\right)e^{i\omega t}\,, 
\end{eqnarray}
where the minus (plus) sign represents bosons (fermions),  

On the other hand, from the bosonic and fermionic correlation functions, Eqs. \eqref{TwoPoint1} and \eqref{TwoPoint2},  the l.h.s. of Eqs. \eqref{FDTB} and \eqref{FDTF} become  
\begin{eqnarray}
G_{\rm Sym}(t)&=&\frac{\alpha' e^{-\frac{k}{r_{h}^{2}}}}{r^{2}_{h}}\int_{0}^{\infty}\frac{d\omega}{\omega}
\left( \frac{4\cos(\omega t)}{e^{\beta\left(\omega-\mu\right)}\mp 1}
+ e^{-i\omega  t}+e^{i\omega  t}\right)\cr
&=&\frac{\alpha' e^{-\frac{k}{r_{h}^{2}}}}{r^{2}_{h}}\int_{0}^{\infty}\frac{d\omega}{|\omega|}
\left( \frac{2\left( e^{-i\omega  t}+e^{i\omega  t}\right)}{e^{\beta\left(|\omega|-\mu\right)}\mp 1}
+ e^{-i\omega t}+e^{i\omega  t}\right)\,,
\label{GF}
\end{eqnarray}
where again the minus (plus) sign represents bosons (fermions). 
Noting that 
\begin{eqnarray}
   \int_{0}^{\infty}
   d\omega f(|\omega|)
   e^{i\omega t}
   +\int_{0}^{\infty}d\omega f(|\omega|)
   e^{-i\omega t}
   =\int_{-\infty}^{\infty}d\omega f(|\omega|)
   e^{i\omega t}\,, 
\end{eqnarray}
we finally have that the symmetric Green's function given by Eq. \eqref{GF} can be rewritten as
\begin{eqnarray}
G_{\rm Sym}
&=&\frac{\alpha' e^{-\frac{k}{r_{h}^{2}}}}{r^{2}_{h}}
\int_{-\infty}^{\infty}\frac{d\omega}{|\omega|}
\left( \frac{2 }{e^{\beta\left(|\omega|-\mu\right)}\mp 1}
+1\right) e^{i\omega  t}\,, 
\end{eqnarray}
which coincides with Eq. \eqref{Vascao}. Then, we checked the Eqs. \eqref{FDTB} and \eqref{FDTF}, completing our verification of the fluctuation-dissipation theorem in the holographic string/gauge deformed metric with bakcreaction at finite temperature and chemical potential. 

\section{Conclusions}\label{conc}

In this work we investigated the effects of a conformal exponential deformation and its backreaction in an AdS-RN spacetime.  Then using the AdS/CFT dictionary we studied the fluctuation and dissipation of a test particle in a thermal bath with finite chemical potential. Solving the EMD field equations we obtain the backreacted horizon function $f(r)$ depending on the chemical potential $\mu$ and the deformation parameter $k$. From this horizon function we obtain the Hawking temperature which is also the temperature of the dual field theory, as a function of the horizon radius $r_h$, as can be seen in Fig. \ref{mukpos}. From the Nambu-Goto action we obtain the equation of motion for the probe string in the metric considered. The solutions of this equation in different regions are obtained from a monodromy patch procedure. Then, we calculate the admittance or linear response, and the friction coefficient $\gamma$. 

An important contribution of this work is given by the admittance calculated in Eq. \eqref{admiT}, which depends on the deformed metric with backreaction at finite temperature and chemical potential. The temperature dependence of this quantity is a non-trivial one. Fig.~\ref{many} summarizes this result in comparison with others discussed previously in the literature. Note that these other models can be understood as particular cases of the one presented here. 

Another important result obtained here is the diffusion of the particle represented by the string endpoint. It was analyzed considering  bosonic and  fermionic statistical distributions. In the two cases we observed that the mean square displacement behaves as $s^{2}\sim \log t$ for large times indicating a sub-diffusive regime. 
Such a behavior was found in ultra-slow scaled Brownian motion, known as Sinai diffusion. Note that Ref. \cite{sinai-like} reports this diffusion process with many physical realizations. 

Therefore the sub-diffusive regime obtained here differs, for instance, from the one of a Brownian motion studied as a dual for the AdS-BTZ black hole presented in Ref.  \cite{deBoer:2008gu}. We interpret this difference due to the presence of a chemical potential in our set up. This assumption is supported by the result presented in Ref. \cite{Caldeira:2020rir}. There, the authors considered a similar gravitational background  in absence of a chemical potential and achieved the typical diffusive behavior  $s^{2}\sim t$. We recovered this result in Sec. \ref{B} for bosons and fermions in the limit $\mu/T\to 0$, in the large time regime.

In the last section we checked the fluctuation dissipation theorem, for both fermionic and bosonic cases:  $\frac{1}{2}\left(\langle x(t)x(0) \rangle+\langle x(0)x(t) \rangle\right) = \mathcal{F}^{-1}\left[\left(1+2n_{B/F}\right)\Im \chi (\omega)\right]$. 
 This was accomplished from the
admittance and the correlation functions calculated in sections \ref{A} and \ref{B}, respectively.

\begin{acknowledgments}

The authors would like to thank Alfonso Ballon Bayona for discussions.  N.G.C. is supported by  Conselho Nacional de Desenvolvimento Científico e Tecnológico (CNPq). H.B.-F. and C.A.D.Z. are partially supported by Conselho Nacional de Desenvolvimento Cient\'{\i}fico e Tecnol\'{o}gico (CNPq) under the grants Nos. 311079/2019-9 and  309982/2018-9, respectively.  
This work is also supported by Coordenação de Aperfeiçoamento de Pessoal de Nível Superior (CAPES),  under finance code 001. 

\end{acknowledgments}

\appendix

\section{Monodromy Patch}\label{apA}

Here in this Appendix we will find approximate analytical solution for the Schr\"{o}dinger-like Eq. equation \eqref{sch} in three different regions in the bulk of the deformed AdS space.

\subsection{Region A: $\omega^{2} \gg V(r)$}
The first region, dubbed as {\bf A}, is nearby the event horizon, $i.e.$ $r\approx r_{h}$. In this case, $V(r)\ll\omega^{2}$, and the Schr\"{o}dinger-like equation reads
\begin{equation}
    \frac{d^{2}\psi(r_{*})}{dr_{*}^{2}}+\omega^{2}\psi(r_{*})=0,
\end{equation}
which has the ingoing solution
\begin{equation}\label{solA}
    \psi(r_{*}) = A_1 e^{-i\omega r_*}.
\end{equation}
Close to the horizon ($r\approx r_{h}$), we can assume that for low frequencies we have $\omega r_* \ll 1$. Then one can expand Eq. \eqref{solA} as: 
\begin{equation}\label{psir*AA}
    \psi(r_{*}) = A_1  -i A_1 \omega r_* .
\end{equation}
Using this equation and the Bogoliubov transformation, $B(r)= -{k}/({2 r^2})-\log (r)$, we can compute $h_{\omega}(r_{*})$ in this region as:
\begin{equation}\label{hAr*}
h_{\omega}^{A}(r_*)=\frac{e^{-\frac{k}{2 r^2_h}}}{r_h}\left(A_{1}- i\omega A_1 r_* \right)\,.
\end{equation}
Close to the horizon, the tortoise  coordinate $r_{*}$ is given by
\begin{eqnarray}
    \int^r \frac{1}{r_{h}^{2}f'(r_{h}) (r'-r_{h})} \, dr'
    = \frac{1}{r_{h}^{2}f'(r_{h})}\log \left(\frac{r}{r_{h}}-1\right)\,. 
\end{eqnarray}
Then, Eq. \eqref{hAr*} gives the expression for $h_\omega^a(r_*)$
\begin{equation}
\label{HA}
   h_{\omega}^{A}(r_*)=A_1\frac{e^{-\frac{k}{2 r^2_h}}}{r_h}\left(1- i\frac{\omega}{r_{h}^{2}f'(r_{h})} \log \left(\frac{r}{r_{h}}-1\right) \right) \,. 
\end{equation}

\subsection{Region B: $V(r)\gg\omega^{2}$}

Now, we analyze the region where $V(r)\gg\omega^{2}$. First, note that the potential term in Eq. \eqref{sch} comes from the term without $\omega^{2}$ in Eq. \eqref{ansatz}. Then, in this second region we can drop this last term and work with \eqref{ansatz} in the form
\begin{equation}
\label{reg2}
    \frac{d }{d r}\left(r^{4}f(r)e^{\frac{k}{r^{2}}}h_{\omega}'\right)=0.
\end{equation}
Solving this equation we obtain
\begin{equation}
    h^{B}_{\omega}(r)=\int^{r}\frac{B_{1}}{r'^{4}f(r')e^{\frac{k}{r'^{2}}}}dr'+B_{2}.
\end{equation}
In the IR regime, we can approximate the integral above using the fact that $f(r)$ has a simple pole at the horizon. Then we can write
\begin{eqnarray}
    h^{B}_{\omega (\rm IR)}(r)
    = \frac{B_{1}}{e^{\frac{k}{r_{h}^{2}}}r_{h}^{4}f'(r_{h})}\log\left(\frac{r}{r_{h}}-1\right)+B_{2}\,. 
\end{eqnarray}
Comparing the last expression with \eqref{HA} we have
\begin{eqnarray}
    B_{1}&=&-i A_{1} r_{h}  \omega  e^{\frac{k}{2 r_{h} ^2}}\,, 
    \cr
    B_{2}&=&A_{1}\frac{e^{-\frac{k}{2r_{h}^{2}}}}{r_{h}}\,.
\end{eqnarray}
For the UV regime, $f(r)\approx 1$ and the solution becomes 
\begin{eqnarray}
   h^{B}_{\omega(\rm UV)}(r)=\int^{r}\frac{B_{1}}{r'^{4}e^{\frac{k}{r'^{2}}}}dr'+B_{2}.
\end{eqnarray}
After integrating we get large values of $r$ 
\begin{equation}
\label{HBUV}
    h^{B}_{\omega(\rm UV)}(r)\approx -\frac{B_{1}}{3 r^3}+B_{2}\,. 
\end{equation}

\subsection{Region C: The Deep UV}
In the deep UV region we have that the horizon function becomes close to the unity and we must solve the equation of motion  in the form
\begin{equation}
    \label{ansatzUV}
    \frac{d }{d r}\left(r^{4}e^{\frac{k}{r^{2}}}h_{\omega}'\right)+\omega^{2}e^{\frac{k}{r^{2}}}h_{\omega}=0\,,
\end{equation}
whose general solution is 
\begin{equation}\label{hwc}
h^C_{\omega}(r)=C_1 \,_1F_1\left(\frac{\omega^{2}}{4k},-\frac{1}{2},-\frac{k}{r^{2}}\right)+ C_2 \frac{k^{3/2}}{r^3}         \, _1F_1\left(\frac{3}{2} + \frac{\omega^{2}}{4k},\frac{5}{2},-\frac{k}{r^{2}}\right)\,. 
\end{equation}

Close to the boundary this solution can be expanded as a power series in $1/r$
\begin{eqnarray}
   h^{C}_{\omega}(r)=C_1+\frac{C_1 \omega ^2}{2 r^2}+\frac{C_2  k ^{3/2}}{r^3}+O\left(\left(\frac{1}{r}\right)^{4}\right) \,. 
\end{eqnarray}
Keeping only terms until $O(\omega)$,we have: 
\begin{eqnarray}
   h^{C}_{\omega}(r)\approx C_1+\frac{C_2  k^{3/2}}{r^3} \,. 
\end{eqnarray}
Matching this result with Eq. \eqref{HBUV}, we find
\begin{eqnarray}
    C_{1}=B_{2}=A_{1}\frac{e^{-\frac{k}{2r_{h}^{2}}}}{r_{h}}, && C_{2}=-\frac{B_1}{3 k ^{3/2}}=\frac{i A_{1} r_{h}  \omega  e^{\frac{k}{2 r_{h} ^2}}}{3 k ^{3/2}} \,. 
\end{eqnarray}

Now we can write the solution close to the boundary as
\begin{eqnarray}
    h^{C}_{\omega}(r)\approx A_1\left(\frac{e^{-\frac{k}{2r_{h}^{2}}}}{r_{h}}
    +\frac{i \omega r_{h} e^{\frac{k}{2 r_{h} ^2}} }{3r^3}\right)\,. 
\end{eqnarray}

\section{Normalization}\label{apB}

The ingoing modes were calculated in the previous section. 
The outgoing modes come from the ingoing modes given by Eq. \eqref{solA}. Then, the outgoing  modes near the horizon read: 
\begin{equation}
    \psi^{out}(r)=A_{2}e^{i\omega r_{*}}\,. 
\end{equation}
At the boundary, analogously to Eq. \eqref{bound}, the outgoing solutions are 
\begin{equation}
    h^{out}_{\omega
    }(r)\approx\left(\frac{e^{-\frac{k}{2r_{h}^{2}}}}{r_{h}}
    -\frac{i \omega r_{h} e^{\frac{k}{2 r_{h} ^2}} }{3r^3}\right)\,. 
\end{equation}
Thus, for large $r$ and small $\omega$ the general solution for the equation of motion can be written as
\begin{equation}
    h_{\omega}^{\rm bound}(r)
    =A\frac{e^{-\frac{k}{2r_{h}^{2}}}}{r_{h}}\left[\left(1
    +\frac{i \omega r^{2}_{h} e^{\frac{k}{ r_{h} ^2}} }{3r^3}\right)+B\left(1
    -\frac{i \omega r^{2}_{h} e^{\frac{k}{ r_{h} ^2}} }{3r^3}\right)\right]\,. 
\end{equation}
Near the boundary we also can write a general solution yet as
\begin{eqnarray}
    h_{\omega}&=&A[h^{out}_{\omega}(r)+Bh^{in}_{\omega}(r)]\nonumber\\
    &=&A\left[ \,_1F_1\left(\frac{\omega ^2}{4 k };-\frac{1}{2};-\frac{k }{r^2}\right)-\frac{i r^{2}_{h}  \omega  e^{\frac{k}{r_{h} ^2}}\,_1F_1\left(\frac{\omega ^2}{4 k }+\frac{3}{2};\frac{5}{2};-\frac{k }{r^2}\right)}{3r^3}\right.\nonumber\\
    & &\left.B\left( \,_1F_1\left(\frac{\omega ^2}{4 k };-\frac{1}{2};-\frac{k }{r^2}\right)+  \frac{i r^{2}_{h}  \omega  e^{\frac{k}{r_{h} ^2}}\,_1F_1\left(\frac{\omega ^2}{4 k }+\frac{3}{2};\frac{5}{2};-\frac{k }{r^2}\right)}{3r^3} \right)  \right]\,. 
\end{eqnarray}
Imposing Neumann boundary condition on the brane at $r=r_{b}$ we can write 
\begin{eqnarray}
\label{BBraneValue}
B&=&- \frac {\Xi}{{\Xi^*}}\,, 
\end{eqnarray}
which is a pure phase, and we define
\begin{eqnarray} 
\Xi &=& i r_{h}^2 e^{\frac{k}{r_{h}^2}} \left[\left(6 k+\omega ^2\right) \, _1F_1\left(\frac{\omega ^2}{4 k}+\frac{5}{2};\frac{7}{2};-\frac{k}{r_{b}^2}\right)-15 r_{b}^2 \, _1F_1\left(\frac{1}{4} \left(\frac{\omega ^2}{k}+6\right);\frac{5}{2};-\frac{k}{r_{b}^2}\right)\right] \nonumber\\
&& + 15 r_{b}^3 \omega  \, _1F_1\left(\frac{\omega ^2}{4 k}+1;\frac{1}{2};-\frac{k}{r_{b}^2}\right)\,. 
\end{eqnarray}

On the other side, close to the horizon we can write the general solution as
\begin{equation}
    h_{\omega}^{\rm hor}(r)
    =A\frac{e^{-\frac{k}{2r^{2}}}}{r}\left[e^{-i \frac{\omega \log \left(\frac{r}{r_{h}}-1\right)}{r_{h}^{2}f'(r_{h})}}+Be^{i \frac{\omega \log \left(\frac{r}{r_{h}}-1\right)}{r_{h}^{2}f'(r_{h})}}\right]\,, 
\end{equation}
and following \cite{deBoer:2008gu}, we impose an extra Neumann boundary condition at $\frac{r}{r_{h}}=1+\epsilon$, $\epsilon\ll 1$. So, we have 
\begin{equation}
    B=-\frac{h_{\omega}^{'in}(r)}{h_{\omega}^{'out}(r)}\Bigg|_{\frac{r}{r_{h}}=1+\epsilon}
\approx e^{-\frac{2 i\omega}{r_{h}^{2}f'(r_{h})}\log\left(\frac{1}{\epsilon}\right)}\,. 
\end{equation}
Matching this equation with \eqref{BBraneValue} one finds that the frequencies are discrete 
\begin{equation}
    \Delta\omega=\frac{\pi r_{h}^{2}f'(r_{h})}{\log\left(\frac{1}{\epsilon}\right)}=\frac{4\pi^{2}T}{\log\left(\frac{1}{\epsilon}\right)}\,. 
\end{equation}
%


From the Klein-Gordon inner product we can normalize the solutions as \cite{deBoer:2008gu,Giataganas:2018ekx, Caldeira:2020sot, wald}
\begin{align}
    (X_{\omega},X_{\omega})
    =\frac{\omega}{\pi \alpha'}\int_{r_{h}}^{r_{b}}dr\;\frac{e^{\frac{k}{r^{2}}}}{f(r)} |h_{\omega}(r)|^{2}=1. \label{IKGP}
\end{align}
This integral is dominated by the near horizon contribution. First,  in this region, at order $\omega$, we have 
\begin{eqnarray}
 |h_{\omega}(r)|^{2}&=&2|h^{out}_{\omega}(r)|^{2}+2\, \Re \left((h^{out }_{\omega}(r))^{2}B^{*}\right)\nonumber\\
 &\approx&2|A|^{2}\frac{e^{-\frac{k}{r^{2}}}}{r^{2}}\left\{1 + \Re \left[B^{*}\left(1+\frac{2i\omega \log \left(\frac{r}{r_{h}}-1\right)}{r_{h}^{2}f'(r_{h})}\right)\right] \right\}
\end{eqnarray}
where $\Re$ represents the real part of the argument. 
Second, from Eq. \eqref{BBraneValue}, up to order $\omega$, we can write that 
\begin{equation}
    B^{*}=1-\frac{r_{b} \omega  \left(2 i e^{-\frac{k}{r_{h}^2}}\right)}{r_{h}^2}+O\left(\frac{\sqrt{|k|}}{r_{b}}\right).
\end{equation}
Note that the limit $\omega \to 0 $ should be taken before the limit $r\to \infty$.

Then, using these results and performing the  integral \eqref{IKGP}, we get in leading order (in $r_{b}$ and $\omega$) close to the horizon at $r=r_{h}+\epsilon r_{h}$
\begin{eqnarray}
\frac{4\omega|A|^{2}}{\pi \alpha'}\int_{r_{h}+\epsilon r_{h}}\frac{dr'}{r^{2}f(r)}\approx\frac{4\omega|A|^{2}}{\pi \alpha'}\int_{r_{h}+\epsilon r_{h}}\frac{dr'}{r_{h}^{2}f'(r_{h})(r'-r_{h})}=\frac{4\omega|A|^{2}}{\pi \alpha'r_{h}^{2}f'(r_{h})}\log\left(\frac{1}{\epsilon}\right)\,. 
\end{eqnarray}
Thus, by imposing the normalization given by Eq.\eqref{IKGP}, we obtain
\begin{equation}
    A=\sqrt{\frac{\pi \alpha'r_{h}^{2}f'(r_{h})}{4\omega \log\left(\frac{1}{\epsilon}\right)}}\,. 
\end{equation}

\end{document}